%% file: ms.tex
\pdfoutput=1
\documentclass[sigconf, nonacm]{acmart}
\usepackage[utf8]{inputenc}
\usepackage{tikz}
\usetikzlibrary{shapes.geometric,decorations.pathmorphing,arrows.meta,calc,patterns}
\input{unicode}

\usepackage[linesnumbered,commentsnumbered,ruled,vlined]{algorithm2e}
\DontPrintSemicolon

\usepackage{xspace}
\usepackage{bm}
\usepackage{enumerate}
\usepackage{changepage}

\usepackage[capitalise,noabbrev]{cleveref}
\newtheorem{theorem}{Theorem}[section]

\newtheorem{lemma}[theorem]{Lemma}

\DeclareMathOperator{\fastrange}{fastrange}

\pgfdeclarepatternformonly{south west lines}{\pgfqpoint{-0pt}{-0pt}}{\pgfqpoint{3pt}{3pt}}{\pgfqpoint{3pt}{3pt}}{
    \pgfsetlinewidth{0.4pt}
    \pgfpathmoveto{\pgfqpoint{0pt}{0pt}}
    \pgfpathlineto{\pgfqpoint{3pt}{3pt}}
    \pgfpathmoveto{\pgfqpoint{2.8pt}{-.2pt}}
    \pgfpathlineto{\pgfqpoint{3.2pt}{.2pt}}
    \pgfpathmoveto{\pgfqpoint{-.2pt}{2.8pt}}
    \pgfpathlineto{\pgfqpoint{.2pt}{3.2pt}}
    \pgfusepath{stroke}}

\def\novldb{1} %
\ifdefined\novldb

\newcommand\vldbpagestyle{plain}
\else

\newcommand\vldbdoi{XX.XX/XXX.XX}
\newcommand\vldbpages{XXX-XXX}
\newcommand\vldbvolume{14}
\newcommand\vldbissue{1}
\newcommand\vldbyear{2020}
\newcommand\vldbauthors{\authors}
\newcommand\vldbtitle{\shorttitle} 
\newcommand\vldbavailabilityurl{http://vldb.org/pvldb/format_vol14.html}
\newcommand\vldbpagestyle{plain} 

\fi

\input{macros}

\begin{document}

\def\footnotetitle{Title inspired by \cite{FAK:CuckooFilterBetter:2013,FAKM:CuckooFilterReallyBetter:2014,GL:XorFilters:2020}.} %
\title[Ribbon filter: practically smaller than Bloom and Xor]{Ribbon filter: practically smaller than Bloom and Xor\ifdefined\footnotetitle\texorpdfstring{${}^1$}{}\fi}

\author{Peter C. Dillinger}
\affiliation{%
  \institution{Facebook, Inc.}
  \streetaddress{Undisclosed}
  \city{Seattle}
  \state{Washington, USA}
  \postcode{98103}
}
\email{peterd@fb.com}

\author{Stefan Walzer}
\affiliation{%
  \institution{University of Cologne}
  \streetaddress{Undisclosed}
  \city{Cologne}
  \country{Germany}
}
\email{walzer@cs.uni-koeln.de}

\begin{abstract}
Filter data structures over-approximate a set of hashable keys, i.e.\ set membership queries may incorrectly come out positive. A filter with \emph{false positive rate} $f ∈ (0,1]$ is known to require $≥ \log₂(1/f)$ bits per key. 
At least for larger $f ≥ 2^{-4}$, existing practical filters require a space overhead of at least 20\% with respect to this information-theoretic bound.

We introduce the Ribbon filter: a new filter for static sets with a broad range of configurable space overheads and false positive rates with competitive speed over that range, especially for larger $f ≥ 2^{-7}$. In many cases, Ribbon is faster than existing filters for the same space overhead, or can achieve space overhead below 10\% with some additional CPU time. An experimental Ribbon design with load balancing can even achieve space overheads below~1\%.

A Ribbon filter resembles an Xor filter modified to maximize locality and is constructed by solving a band-like linear system over Boolean variables. In previous work, Dietzfelbinger and Walzer describe this linear system and an efficient Gaussian solver. We present and analyze a faster, more adaptable solving process we call “\textbf{R}apid \textbf{I}ncremental \textbf{B}oolean \textbf{B}anding \textbf{ON} the fly,” which resembles hash table construction. We also present and analyze an attractive Ribbon variant based on making the linear system homogeneous, and describe several more practical enhancements.
\end{abstract}

\maketitle

\ifdefined\novldb
\enlargethispage{-5em} %

\begingroup
\pagestyle{\vldbpagestyle}
\renewcommand\thefootnote{}\footnote{\noindent
This work is licensed under the Creative Commons BY-NC-ND 4.0 International License. Visit \url{https://creativecommons.org/licenses/by-nc-nd/4.0/} to view a copy of this license. Copyright is held by the owner/author(s).
}\addtocounter{footnote}{-1}\endgroup

\else

\pagestyle{\vldbpagestyle}
\begingroup\small\noindent\raggedright\textbf{PVLDB Reference Format:}\\
\vldbauthors. \vldbtitle. PVLDB, \vldbvolume(\vldbissue): \vldbpages, \vldbyear.\\
\href{https://doi.org/\vldbdoi}{doi:\vldbdoi}
\endgroup
\begingroup
\renewcommand\thefootnote{}\footnote{\noindent
This work is licensed under the Creative Commons BY-NC-ND 4.0 International License. Visit \url{https://creativecommons.org/licenses/by-nc-nd/4.0/} to view a copy of this license. For any use beyond those covered by this license, obtain permission by emailing \href{mailto:info@vldb.org}{info@vldb.org}. Copyright is held by the owner/author(s). Publication rights licensed to the VLDB Endowment. \\
\raggedright Proceedings of the VLDB Endowment, Vol. \vldbvolume, No. \vldbissue\ %
ISSN 2150-8097. \\
\href{https://doi.org/\vldbdoi}{doi:\vldbdoi} \\
}\addtocounter{footnote}{-1}\endgroup

\ifdefempty{\vldbavailabilityurl}{}{
\vspace{.3cm}
\begingroup\small\noindent\raggedright\textbf{PVLDB Artifact Availability:}\\
The source code, data, and/or other artifacts have been made available at \url{\vldbavailabilityurl}.
\endgroup
}

\fi

\ifdefined\footnotetitle
    \stepcounter{footnote}
    \footnotetext{\footnotetitle}
\fi

\input{intro-v4}

\input{from-retrieval-to-filters}

\input{theory}

\section{Homogeneous Ribbon Filters}
\input{homog}

\section{Making Ribbon Practical}
\input{configurability}

\input{solution-layout}

\input{practical-hashing}

\input{standard-scalability}

\input{balanced-ribbon}

\input{experiments}

\input{conc-future}

\begin{acks}
 We thank Martin Dietzfelbinger for early contributions to this line of research, and Peter Sanders for later discussions. We thank others at Facebook for their supporting roles, including Jay Zhuang, Siying Dong, Shrikanth Shankar, Affan Dar, and Ramkumar Vadivelu.
\end{acks}

\bibliographystyle{ACM-Reference-Format}
\bibliography{bibliographie}

\end{document}

%% file: unicode.tex
\usepackage{newunicodechar}
\usepackage{silence}
\newunicodechar{♢}{\tikz \node[inner sep=1.5,draw,diamond] {};}
\newunicodechar{☆}{\tikz \node[inner sep=1,draw,star,star point ratio=2] {};}
\newunicodechar{△}{\kern 0.5pt\tikz \node[inner sep=1.2,draw,regular polygon,regular polygon sides=3] {};\kern 0.5pt}
\newunicodechar{⬜}{\kern 0.5pt\tikz \node[inner sep=1.7,draw,regular polygon,regular polygon sides=4] {};\kern 0.5pt}
\newunicodechar{◯}{\tikz[baseline=-3pt] \node[inner sep=1.7,draw,cloud,cloud puffs=4,cloud puff arc=190] {};}

\newunicodechar{⊥}{\bot}
\newunicodechar{•}{\item}
\newunicodechar{✓}{\checkmark}
\newunicodechar{✗}{\xmark}
\newunicodechar{…}{\dots}
\newunicodechar{≔}{\coloneqq}
\newunicodechar{⁻}{^-}
\newunicodechar{⁺}{^+}
\newunicodechar{₋}{_-}
\newunicodechar{₊}{_+}
\newunicodechar{ℓ}{\ell}
\newunicodechar{•}{\item}
\newunicodechar{…}{\dots}
\newunicodechar{≔}{\coloneqq}
\newunicodechar{≤}{\leq}
\newunicodechar{≥}{\geq}
\newunicodechar{≰}{\nleq}
\newunicodechar{≱}{\ngeq}
\newunicodechar{⊕}{\oplus}
\newunicodechar{⊗}{\otimes}
\newunicodechar{≠}{\neq}
\newunicodechar{¬}{\neg}
\newunicodechar{≡}{\equiv}
\newunicodechar{₀}{_0}
\newunicodechar{₁}{_1}
\newunicodechar{₂}{_2}
\newunicodechar{₃}{_3}
\newunicodechar{₄}{_4}
\newunicodechar{₅}{_5}
\newunicodechar{₆}{_6}
\newunicodechar{₇}{_7}
\newunicodechar{₈}{_8}
\newunicodechar{₉}{_9}
\newunicodechar{⁰}{^0}
\newunicodechar{¹}{^1}
\newunicodechar{²}{^2}
\newunicodechar{³}{^3}
\newunicodechar{⁴}{^4}
\newunicodechar{⁵}{^5}
\newunicodechar{⁶}{^6}
\newunicodechar{⁷}{^7}
\newunicodechar{⁸}{^8}
\newunicodechar{⁹}{^9}
\newunicodechar{∈}{\in}
\newunicodechar{∉}{\notin}
\newunicodechar{⊂}{\subset}
\newunicodechar{⊃}{\supset}
\newunicodechar{⊆}{\subseteq}
\newunicodechar{⊇}{\supseteq}
\newunicodechar{⊄}{\nsubset}
\newunicodechar{⊅}{\nsupset}
\newunicodechar{⊈}{\nsubseteq}
\newunicodechar{⊉}{\nsupseteq}
\newunicodechar{∪}{\cup}
\newunicodechar{∩}{\cap}
\newunicodechar{∀}{\forall}
\newunicodechar{∃}{\exists}
\newunicodechar{∄}{\nexists}
\newunicodechar{∨}{\vee}
\newunicodechar{∧}{\wedge}
\newunicodechar{ℝ}{\mathbb{R}}
\newunicodechar{ℕ}{\mathbb{N}}
\newunicodechar{ℤ}{\mathbb{Z}}
\newunicodechar{·}{\cdot}
\newunicodechar{∘}{\circ}
\newunicodechar{×}{\times}
\newunicodechar{↑}{\uparrow}
\newunicodechar{↓}{\downarrow}
\newunicodechar{→}{\rightarrow}
\newunicodechar{←}{\leftarrow}
\newunicodechar{⇒}{\Rightarrow}
\newunicodechar{⇐}{\Leftarrow}
\newunicodechar{↔}{\leftrightarrow}
\newunicodechar{⇔}{\Leftrightarrow}
\newunicodechar{↦}{\mapsto}
\newunicodechar{∅}{\varnothing}
\newunicodechar{∞}{\infty}
\newunicodechar{≅}{\cong}
\newunicodechar{≈}{\approx}
\newunicodechar{ℓ}{\ell}
\newunicodechar{⌊}{\lfloor}
\newunicodechar{⌋}{\rfloor}
\newunicodechar{⌈}{\lceil}
\newunicodechar{⌉}{\rceil}

\newunicodechar{α}{\alpha}
\newunicodechar{β}{\beta}
\newunicodechar{γ}{\gamma}
\newunicodechar{Γ}{\Gamma}
\newunicodechar{δ}{\delta}
\newunicodechar{Δ}{\Delta}
\newunicodechar{ε}{\varepsilon}
\newunicodechar{ζ}{\zeta}
\newunicodechar{η}{\eta}
\newunicodechar{θ}{\theta}
\newunicodechar{Θ}{\Theta}
\newunicodechar{ι}{\iota}
\newunicodechar{κ}{\kappa}
\newunicodechar{λ}{\lambda}
\newunicodechar{Λ}{\Lambda}
\newunicodechar{μ}{\mu}
\newunicodechar{ν}{\nu}
\newunicodechar{ξ}{\xi}
\newunicodechar{Ξ}{\Xi}
\newunicodechar{π}{\pi}
\newunicodechar{Π}{\Pi}
\newunicodechar{ρ}{\rho}
\newunicodechar{σ}{\sigma}
\newunicodechar{Σ}{\Sigma}
\newunicodechar{τ}{\tau}
\newunicodechar{υ}{\upsilon}
\newunicodechar{ϒ}{\Upsilon}
\newunicodechar{φ}{\phi}
\newunicodechar{ϕ}{\varphi}
\newunicodechar{Φ}{\Phi}
\newunicodechar{χ}{\chi}
\newunicodechar{ψ}{\psi}
\newunicodechar{Ψ}{\Psi}
\newunicodechar{ω}{\omega}
\newunicodechar{Ω}{\Omega}

%% file: macros.tex
\def\U{\mathcal{U}}
\def\O{\mathcal{O}}
\def\sgauss{\textsc{sgauss}\xspace}
\def\ribbon{Ribbon\xspace} %
\def\E{\mathbb{E}}
\def\fpr{\ensuremath{f}}
\def\refrel#1#2#3{\stackrel{\text{#2 \ref{#1}}}{#3}}
\def\myparagraph#1{\smallskip\noindent\textbf{#1}}

%% file: intro-v4.tex
\section{Introduction}
\label{sec:intro}
\def\true{\textsc{true}\xspace}
\def\false{\textsc{false}\xspace}

\myparagraph{Background and motivation.}
The primary motivation for this work is optimizing data retrieval, especially in systems aggregating immutable data resources. In the example of LSM-tree storage~\cite{OCGO:LSM:1996}, persisted key-value data is primarily split among immutable data files. A crucial strategy for reducing I/O in key look-ups is filtering accesses using an in-memory data structure. In a common configuration, each data file has an associated Bloom filter~\cite{B:Space:1970} representing the set of keys with associated data in that file. The Bloom filter has some false positive (FP) rate, which is the probability that querying a key not added returns \true (positive). For example, configuring the Bloom filter to use 10 bits of space per added key yields an FP rate of just under 1\%, regardless of the size or structure of the keys themselves\footnote{Learned filters~\cite{VKKM:Learned:2020} or tries~\cite{ZLLAKKP:SuRF:2018} can take advantage of regularities in the key set. A space-efficient hash table~\cite{Cleary:Tables:1984} can take advantage of a densely covered key space. We focus on the general case.}. Thus, the Bloom filter \emph{filters out} almost all specific key queries\footnote{Static filters can also support range queries, either through \emph{prefix Bloom}~\cite{MDL:MyRocks:2020} or more sophisticated schemes~\cite{LCKDQI:Rosetta:2020}.} to data files that would find no relevant data. False negative (FN) queries would be incorrect for this application and must never occur.

Blocked Bloom filters~\cite{Putze:Efficient-Bloom-Filters:2009,LNKB:Bloom:2019} are a popular Bloom variant because they are extremely fast. We do not expect to improve upon this solution\footnotemark\ for short-lived applications such as database joins or the smallest levels of an LSM-tree. However, Bloom filters use at least 44\% more space (``space overhead'') than the information-theoretic lower bound of $λ = \log₂ (1/f)$ bits per key for a hashed filter with FP rate $f$~\cite[Section 2.2]{BM:Survey:2003}. Blocked Bloom can exceed 50\% space overhead for small $f$.
\footnotetext{In \cref{sec:experiments} we mention and use another blocked Bloom implementation with new trade-offs~\cite{Dillinger:RocksDBBloom}.}

\begin{figure}[htb]
    \includegraphics[page=1,width=\linewidth]{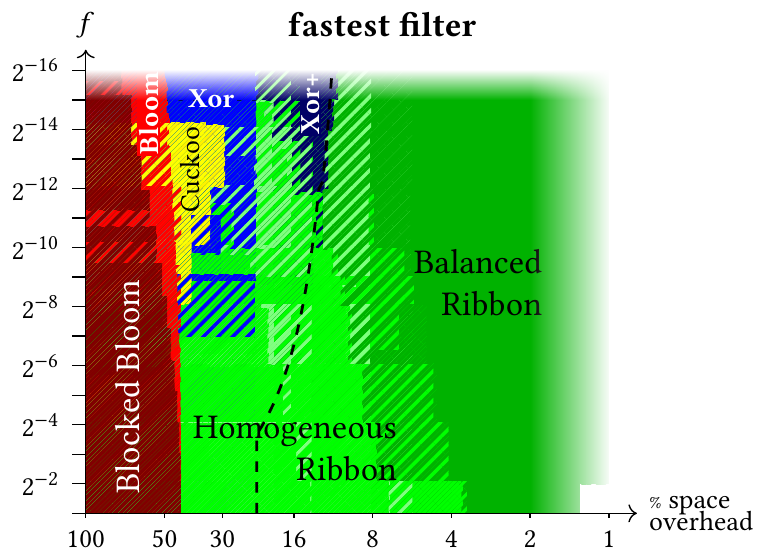}
    \includegraphics[page=2,width=0.5\linewidth]{BenchmarkFigure.pdf}\includegraphics[page=3,width=0.5\linewidth]{BenchmarkFigure.pdf}
    \caption[fragile]{
        \textbf{(a)}
        Fastest filter with the given combination of space overhead and false positive rate, considering a mix of construction and query times.
        \textbf{(b)}
        Construction and query times for fastest approaches in \textbf{(a)}.
    }
    \label{fig:performance-comparison}
\end{figure}

In this work we focus on saving space in static filters, and optimizing the CPU time required for saving space. Our presentation and validation are kept general, but an intended application is optimizing accesses to the largest levels of an LSM-tree, where it should be worth CPU time to save space in long-lived memory-resident structures\footnote{In some large-scale applications using  RocksDB~\cite{DKJS:RocksDB:2021} for LSM-tree storage, we observe roughly 10\% of memory and roughly 1\% of CPU used in blocked Bloom filters. The size-weighted average age of a live filter is about three days. An experimental Ribbon filter option was added to RocksDB in version 6.15.0 (November 2020).}. In~\cite{DAI:Monkey:2017} it is shown that a relatively high FP rate for these levels is best for overall efficiency. However, the space savings offered by existing practical Bloom filter alternatives is limited (dotted line in \cref{fig:performance-comparison} \textbf{(a)}), especially for higher FP rates, $λ ≤ 5$.

\ifdefined\novldb
\enlargethispage{-5em} %
\fi

\myparagraph{Bloom filter and alternatives.}
We can categorize hashed filters by the logical structure of a query:
\begin{description}
    •[OR probing:] Cuckoo filters~\cite{FAK:CuckooFilterBetter:2013,FAKM:CuckooFilterReallyBetter:2014,Eppstein:CuckooFilter:2016,BJ:MortonFilters:2020}, Quotient filters~\cite{Cleary:Tables:1984,PPRS:BloomReplacement:2005,DM:Storage:2009,BFJKKMMSS:QuotientFilters:2012}, and most others return \true for a query if any one of several probed locations is a hashed data match, much like hash table look-ups. This design is great for supporting dynamic add and delete, but all known instances use $(1 + ε)λ + μ$ bits per key where $μ > 1.44$, or often $μ = 3$ for speed\footnotemark. Even with $ε \approx 0.05$, the space overhead is large for small $λ$.
    \footnotetext{$μ > 1.44$ can be explained by these structures approximating a minimal perfect hash~\cite{FK:On_the_Size:1984}, a near-strict ordering of keys based on the location of their matching entry in the structure, on top of $λ$ payload bits per key. This is an observation about existing structures, not necessarily a fundamental limitation.}
    •[AND probing:] A Bloom filter query returns \true iff all probed locations (bits) are a match (set to 1).%
    •[XOR probing:] Xor filters~\cite{GL:XorFilters:2020,DP:Succinct:2008,Vigna:Fast-Scalable-Construction-of-Functions:2016,CKRT:The_Bloomier:2004,BPZ:Practical:2013} return \true for a query iff the bitwise exclusive-or (XOR) of all probed locations is a hashed data match. XOR probing is only known to work with static filters.
\end{description}

Structures using XOR probing are the most promising for space efficient static filters. They are constructed by solving a linear system with one constraint per key ensuring that querying the key returns \true. Standard Xor filters use a fast solving process called \emph{peeling} that limits their space efficiency to $≥ 1.22λ$ bits per key\footnotemark, though a variant with fast compression, the Xor+ filter~\cite{GL:XorFilters:2020}, uses roughly $1.08λ + 0.5$ bits per key, which is an improvement for larger $λ$. Using structured Gaussian elimination instead of peeling~\cite{Vigna:Fast-Scalable-Construction-of-Functions:2016,DW:Retrieval-log-extra-bits:2019} offers better space efficiency, but construction times are considered impractical for many applications.

\footnotetext{A new spatially-coupled construction for Xor filters~\cite{W:SpatialCoupling:2021} promises lower space overheads and in some cases slightly improved construction time with peeling. Simulations indicate a number of keys on the order of $10^6$ or more is needed for most of the benefit, limiting the known generality of the approach. Configuration in practice is not well understood.}

\myparagraph{Core contribution.} We introduce a faster, simplified, and more adaptable Gaussian elimination algorithm (\ribbon) for the static function data structure from~\cite{DW:One-Block-per-Row:2019}. Based on Ribbon, we develop a family of practical and highly space-efficient XOR-probed filters.

\myparagraph{Results and comparison.}
\cref{fig:performance-comparison} \textbf{(a)} summarizes extensive benchmarking data by indicating which structure is fastest for satisfying various space and FP rate requirements for a static filter. For “fastest” we consider the sum of the construction time per key and three query times (measured for $x ∈ S$, $x ∉ S$ and a mixed data set).\footnote{The “right” weighing of construction and query time clearly depends on the use case. Because LSM-trees are especially useful for write-heavy workloads requiring good read latency, we find this a reasonable ratio for at least that use case. If a 4KB filter memory page has a lifetime as long as two weeks and at least one negative (useful) query per added key ($n$ roughly $2^{12}$) is seen, that satisfies the current rule of five minutes for caching SSD storage in RAM~\cite{AGBA:FiveMinuteRule:2019}.}

Although we compare Ribbon with many approaches implemented in the fastfilter benchmark library~\cite{Lemire:fastfilter:2020}, only variants of Bloom, Cuckoo, Xor, and Ribbon emerge as winners. Specifically, the color at point $(x,y)$ indicates the fastest filter with space overhead at most $x$ and FP rate $f ∈ [y/2,y·2]$ for $n = 10⁷$ keys. Diagonal shading indicates different winners for $n = 10⁶$ and $n = 10⁸$. 
The timings for the winning approach are also shown in \cref{fig:performance-comparison} \textbf{(b)}.
We observe the following.
\begin{itemize}
    • Ribbon wins to the right of the dotted line because none of the competing approaches achieve space overhead this low.
    • Ribbon wins in some territory previously occupied by Xor and Xor+ filters, mostly for $f > 2^{-8}$ ($λ < 8$) from Xor and $f > 2^{-12}$ ($λ < 12$) for Xor+. %
    In these cases, Ribbon-style Gaussian elimination is faster than peeling.
    • Blocked Bloom filters are still the fastest whenever applicable, though Cuckoo and Xor take some of the remaining >44\% territory (and nearby) for small FP rates.
\end{itemize}

\myparagraph{Outline.} The paper is structured as follows.
\begin{description}
    •[Section 2.] We briefly review data structures for \textbf{static functions} and how they give rise to filters.
    •[Section 3.] We describe and analyze the new \ribbon construction algorithm, which preserves asymptotic guarantees from \cite{DW:One-Block-per-Row:2019}. We also show how to improve the space efficiency of small Ribbon structures (``smash''). These features go into the Standard Ribbon filter, which in practice has increasing space overhead or running time as the number of keys increases.\footnote{If the processor word size is assumed to be $Ω(\log n)$, query time and space overhead can be kept constant. We make no such assumption here.}
    •[Section 4.] We present the Homogeneous Ribbon filter, which shares many desirable properties with blocked Bloom filters: construction success is guaranteed, and scaling to any number of keys is efficient. Homogeneous Ribbon does not build on static functions in the standard way, which simplifies implementation but complicates analysis.
    •[Section 5.] We describe some practical enhancements and issues for Ribbon filters, including (1) efficiently utilizing any amount of memory for any number of keys, (2) laying out data for efficient queries, (3) efficiently satisfying hashing requirements, and (4) scaling Standard Ribbon with data sharding.
    •[Section 6.] We present Balanced Ribbon, an experimental extension of Standard Ribbon that uses a greedy load balancing scheme within a contiguous ribbon. This scales the extreme space efficiency of small Standard Ribbon filters to very large $n$, such as only $1.005λ + 0.008$ bits per key with practical construction and query times.
    •[Section 7.] We present more experimental validation.
\end{description}

%% file: from-retrieval-to-filters.tex
\section{From Static Functions to Filters}

\def\svec#1{\smash{\vec{#1}}}

\myparagraph{Approximate membership queries and filters.}
An approximate membership query filter – \emph{filter} for short – represents a set $S ⊆ \U$ from some universe $\U$. A membership query with $x ∈ S$ must return \textsc{true}, while a query with $x ∈ \U \setminus S$ may return \textsc{true} with probability at most $\fpr$ where $\fpr > 0$ is the \emph{false positive (FP) rate}.

\myparagraph{Static functions.}
A \emph{static function} is a data structure (sometimes called “retrieval data structure”) representing a function $b: S → \{0,1\}^r$ for some set $S ⊆ \U$ of keys. A query for $x ∈ S$ must return $b(x)$ but a query for $x ∈ \U \setminus S$ may return any value from $\{0,1\}^r$. Membership queries (“is $x ∈ S$?”) are not supported.

\myparagraph{Static Functions from Linear Systems.}
A well-known way for constructing static functions \cite{DP:Succinct:2008,Vigna:Fast-Scalable-Construction-of-Functions:2016,ADR:Experimental:2009,P:An_Optimal:2009,CKRT:The_Bloomier:2004,BPZ:Practical:2013} uses a hash function to associate each key $x ∈ \U$ with a set $h(x) ⊆ [m]$ for some $m ≥ n = |S|$. By $\svec{h}(x) ∈ \{0,1\}^m$ we denote the characteristic (row) vector of $h(x)$. If $(\svec{h}(x))_{x ∈ S}$ are linearly independent in the vector space $\{0,1\}^m$ over the two-element field then the system $(\svec{h}(x)·Z = b(x))_{x∈S}$ of linear equations has a solution $Z ∈ \{0,1\}^{m × r}$. The static function is then given by $h$ and $Z$. Most memory is used for the $mr$ bits of $Z$, which takes $\frac{m}{n}r$ bits per key. A query for $x ∈ \U$ returns
\begin{equation}
    \mathrm{query}(x) := \svec{h}(x)·Z = \bigoplus_{i ∈ h(x)} Z_i.\label{eq:xor-filter-query}
\end{equation}
where $Z_i$ denotes the $i$-th row of $Z$. Since queries involve $|h(x)|·r$ bits from $Z$ fast query times require sparse $\svec{h}(x)$.

Several constructions choose $h$ such that $\svec{h}(x)$ contains exactly three $1$-bits in random positions \cite{BPZ:Practical:2013,Vigna:Fast-Scalable-Construction-of-Functions:2016}. In this case $\frac{n}{m}$ must not exceed the corresponding XORSAT threshold \cite{PS:The_Satisfiability:2016,DGMMPR:Tight:2010} $c₃^* ≈ 0.92$. If a greedy algorithm is used for solving the linear system, then $\frac{n}{m}$ must not exceed the \emph{peeling threshold} $\smash{c^{Δ}₃} ≈ 0.82$.

The space usage is roughly $(1+ε)r$ bits per key when $m = (1+ε)n$. The first paper to achieve $ε = o(1)$ is \cite{P:An_Optimal:2009}. Even $ε = \O(\log n / n)$ is possible, albeit with mediocre construction time \cite{DW:Retrieval-log-extra-bits:2019}. A recent more practical contribution that (more humbly) aims for small \emph{constant} $ε > 0$ \cite{DW:One-Block-per-Row:2019} will be the starting point of our own construction.

\myparagraph{Xor filters.}
There is a straightforward way to obtain a filter with FP rate $2^{-r}$ from an $r$-bit static function as pointed out in \cite[Observation 1]{DP:Succinct:2008}.
Simply pick a random fingerprint function $b: \U → \{0,1\}^r$ (a hash function) and store its restriction $b_S: S → \{0,1\}^r$ as a static function. Then for any $x ∈ S$ the static function reproduces $b_S(x) = b(x)$ while for $x ∈ \U \setminus S$ the returned value will match $b(x)$ only with probability $2^{-r}$ (because $b(x)$ is random and plays no role in the construction of the static function).

Such filters inherit the performance of the underlying static function, giving them the potential to be “Faster and Smaller Than Bloom and Cuckoo Filters” as claimed in \cite{GL:XorFilters:2020}, when dynamic insertions and deletions are not required.
A standard construction with $h(x)$ being a fully random set of size $3$ is appropriately named \emph{Xor filter}~\cite{GL:XorFilters:2020}.

%% file: theory.tex
\section{Ribbon Retrieval and Ribbon Filters}

By enriching the \sgauss static function\footnote{Strictly speaking, \sgauss is the name of the construction algorithm of the otherwise unnamed data structure.} from \cite{DW:One-Block-per-Row:2019}, we obtain the \emph{\ribbon static function} which can be used as a \emph{\ribbon filter}.
Since Ribbon filters also retrieve fingerprints using \cref{eq:xor-filter-query}—just with a different choice of $h$—they can be seen as (non-standard) Xor filters.

\myparagraph{The \sgauss construction.} For a parameter $w ∈ ℕ$ that we call the \emph{ribbon width}, the vector $\svec{h}(x) ∈ \{0,1\}^m$ is given by a random \emph{starting position} $s(x) ∈ [m-w-1]$ and a random \emph{coefficient vector} $c(x) ∈ \{0,1\}^w$ as $\svec{h}(x) = 0^{s-1}c(x)0^{m-s-w+1}$. Note that even though $m$-bit vectors like $\svec{h}(x)$ are used to simplify mathematical discussion, such vectors can be represented using $\log(m)+w$ bits.

\begin{figure}[tb]
    \centering
    \includegraphics[page=1]{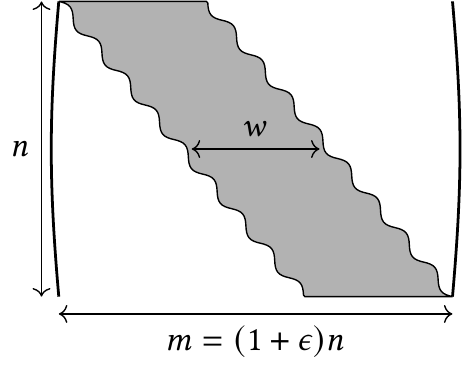}
    \caption[fragile]{Typical shape of the random matrix with rows $\bm{(\vec{h}(x))_{x ∈ S}}$ sorted by starting positions. The shaded “ribbon” region contains random bits. Gaussian elimination never causes any fill-in outside of the ribbon.}
    \label{fig:sgauss-matrix}
\end{figure}

The matrix with rows $(\svec{h}(x))_{x ∈ S}$ sorted by $s(x)$ has all of its $1$-entries in a “ribbon” of width $w$ that randomly passes through the matrix from the top left to the bottom right, as in \cref{fig:sgauss-matrix}.
The authors of \cite{DW:One-Block-per-Row:2019} showed that a solution $Z ∈ \{0,1\}^{m × r}$ to $(\svec{h}(x)·Z = b(x))_{x ∈ S}$ can be computed quickly:

\begin{theorem}[{\cite[Thm 2]{DW:One-Block-per-Row:2019}}]
    \label{thm:sgauss}
    For any constant $0 < ε < \frac{1}{2}$, $w = \smash{\frac{\log n}{ε}}$ and $\frac{n}{m} = 1-ε$, with high probability the linear system $(\svec{h}(x)·Z = b(x))_{x ∈ S}$ is solvable for any $r ∈ ℕ$ and any $b : S → \{0,1\}^r$. Moreover, after sorting $(\svec{h}(x))_{x ∈ S}$ by $s(x)$, Gaussian elimination can compute a solution $Z$ in expected time $\O(n/ε²)$.
\end{theorem}

\myparagraph{Boolean banding on the fly.} For \ribbon we start with the same hash function $\svec{h}$ as in \sgauss. For slightly improved presentation, execution speed, and chance of construction success, we force coefficient vectors $c(x)$ to start with $1$.%
\footnote{In asymptotic considerations this change is inconsequential (and mildly annoying). For better alignment with \cite{DW:One-Block-per-Row:2019} our theorems still assume that $c(x)$ is uniformly distributed in $\{0,1\}^w$.\label{fn:no-leading-1-in-theory}}

\begin{figure}
    \centering
    \includegraphics[page=4]{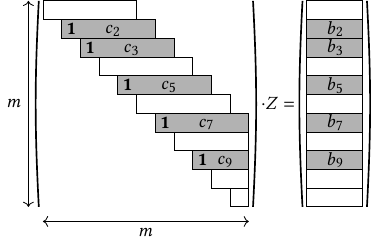}
    \caption{Shape of the linear system $\bm{M}$ central to Boolean banding on the fly.}
    \label{fig:ribbon-matrix}
\end{figure}

The main difference lies in how we solve the linear system. The \emph{insertion phase} maintains a reduced system $M$ of linear equations using on-the-fly Gaussian elimination~\cite{BGV:OnTheFlyGauss:2010}. This system is of the form shown in \cref{fig:ribbon-matrix} and has $m$ rows. 
Each row $i$ is represented by a $w$-bit vector $c_i ∈ \{0,1\}^w$ and $b_i ∈ \{0,1\}^r$. Logically, the $i$-th row is either empty ($c_i = 0^w$) or specifies a linear equation $c_i · Z_{[i,i+w)} = b_i$ where $c_i$ starts with a $1$.
With $Z_{[i,i+w)} ∈ \{0,1\}^{w × r}$ we refer to rows $i,…,i+w-1$ of $Z$. We ensure $c_i · Z_{[i,i+w)}$ is well-defined even when $i + w - 1 > m$ with the invariant that $c_i$ values never select ``out of bounds'' rows of $Z$.

We consider the equations $(\svec{h}(x)·Z = b(x))_{x ∈ S}$ one by one, in arbitrary order, and try to integrate each into $M$ using \cref{algo:ribbon-insertion}, which we explain now.
A key's equation may be modified several times before it can be added to $M$, but a loop invariant is that its form is
\begin{equation}
    c · Z_{[i,i+w)} = b \text{ for $i ∈ [m]$, $c ∈ 1∘\{0,1\}^{w-1}$, $b ∈ \{0,1\}^r$.}\label{eq:form-of-equation}
\end{equation}
The initial equation $\svec{h}(x)·Z = b(x)$ of key $x ∈ S$ has this form with $i = s(x)$, $c = c(x)$ and $b = b(x)$. We proceed it as follows.
\begin{description}
    •[Case 1:] In the simplest case, row $i$ of $M$ is empty and we can incorporate \cref{eq:form-of-equation} as the new $i$-th row of $M$.
    •[Case 2:] Otherwise row $i$ of $M$ is already occupied by an equation $c_i · Z_{[i,i+w)} = b_i$. We compute the summed equation
    \begin{equation}
        c' · Z_{[i,i+w)} = b' \text{ with $c' = c ⊕ c_i$ and $b' = b ⊕ b_i$,}\label{eq:form-of-eq-modified}
    \end{equation}
    which, in the presence of row $i$ of $M$, puts the same constraint on $Z$ as \cref{eq:form-of-equation}.
    Both $c$ and $c_i$ start with $1$, so $c'$ starts with $0$. We consider the following sub-cases. 
    \begin{description}
        •[Case 2.1:] $c' = 0^w$ and $b' = 0^r$.
        The equation is void and can be ignored. This case is reached when the key's original equation is implied by equations previously added to $M$.
        •[Case 2.2:] $c' = 0^w$ and $b' ≠ 0^r$.
        The equation is unsatisfiable. This case is reached when the key's original equation is inconsistent with equations previously added to $M$.
        •[Case 2.3:] $c'$ starts with $j > 0$ zeroes followed by a $1$. Then \cref{eq:form-of-eq-modified} can be rewritten as $c'' · Z_{[i',i'+w)} = b'$ where $i' = i + j$ and $c''$ is obtained from $c'$ by discarding the $j$ leading zeroes of $c$ and appending $j$ trailing zeroes.\\
        Note that in the bit-shift of \cref{algo:ribbon-insertion} the roles of “leading” and “trailing” may seem reversed because the least-significant “first” bit of a word is conventionally thought of as the “right-most” bit.
    \end{description}
\end{description}
Termination is guaranteed since $i$ increases with each loop iteration.

\SetKwFor{Loop}{loop}{}{again}
\SetKwData{shift}{shift}
\SetKwFunction{ctz}{ctz}
\begin{algorithm}
$i ← s(x)$\;
$c ← c(x)$\;

$b ← b(x)$\;
 \Loop{}{
  \If(\tcp*[h]{row $i$ of $M$ is empty}){$M.c[i] = 0$}{
   $M.c[i] ← c$\;
   $M.b[i] ← b$\;
   \Return \textsc{success} (inserted)
  }
  $c ← c ⊕ M.c[i]$\;
  $b ← b ⊕ M.b[i]$\;
  \If{$c = 0$}{
   \lIf{$b = 0$}{ \Return \textsc{success} (redundant) }
   \lElse{ \Return \textsc{failure} (inconsistent) }
  }
  $j ← \mathrm{findFirstSet}(c)$ \tcp{a.k.a. BitScanForward}
  $i ← i + j$\;
  $c ← c >> j$ \tcp{logical shift last toward first}
 }
 \caption[fragile]{\hbox{Adding a key's equation to the linear system $M$.}}
 \label{algo:ribbon-insertion}
\end{algorithm}

Once equations for all keys are successfully inserted, we obtain a solution $Z$ to $M$ in the \emph{back substitution phase}. The rows of $Z$ are obtained from bottom to top. If row $i$ of $M$ contains an equation then this equation uniquely determines row $i$ of $Z$ in terms of later rows of $Z$. If row $i$ of $M$ is empty, then row $i$ of $Z$ can be initialized arbitrarily.

\myparagraph{“On-the-fly” and “incremental.”} The insertion phase of ribbon is on-the-fly in the sense that for a sequence $S = (x₁,x₂,x₃,…)$ of keys we can easily determine the longest prefix $(x₁,…,x_n)$ of $S$ for which construction succeeds: Simply insert keys until the first failure. The insertion phase is incremental because we can easily undo a set of most recent successful insertions: Simply remove the rows from $M$ that were added. These properties are not shared by \sgauss and will be exploited in \cref{sec:balanced}.

\myparagraph{Analysis.} All performance guarantees for the construction algorithm carry over from \sgauss as follows.

\begin{theorem}
    Let $S ⊆ \U$ be an arbitrary key set.
    \begin{enumerate}[(i)]
        • If an \sgauss construction succeeds for $S$ then so does the \ribbon construction. %
        • If both constructions succeed on $S$, expected running times coincide up to constant factors.
    \end{enumerate}
\end{theorem}

\vspace{-\baselineskip}\begin{proof}\ 
\begin{enumerate}[(i)]
    • This is unsurprising as both approaches attempt to solve the same linear system.\footnote{See \cref{fn:no-leading-1-in-theory}.} A superficial difference concerns redundant equations. In \cref{algo:ribbon-insertion} it is natural to ignore them. \sgauss treats them as failures, to avoid special cases during back-substitution.
    • Consider a set $S$ on which \sgauss succeeds, i.e. $S$ gives rise to a solvable system without redundant equations. The back-substitution phases are identical in both algorithms. The analysis of the \ribbon insertion phase hinges on counting row additions. Each key $x ∈ S$ has a starting position $s(x)$ and causes some row $i(x)$ of $M$ to be filled. The number of row additions for the insertion is clearly at most the \emph{displacement} $i(x) - s(x)$ of $x$.
    Summing over all keys yields
    \[ D = \sum_{x ∈ S} i(x) - s(x) = \sum_{i ∈ P} i - \sum_{x ∈ S} s(x) \]
    where $P = \{i(x) \mid x ∈ S\}$ is the set of row-indices of $M$ that end up being occupied.
    A crucial observation is that even though the values $i(x)$ depend on the insertion order of the keys, the set $P$ does not. Indeed, for any $j ∈ [m]$ the value $|P ∩ \{1,…j\}|$ is the rank of the sub-matrix of $M$ formed by its first $j$ columns, which is invariant under row operations. So no matter in what order the keys of $S$ are inserted, we always observe the same sets $P ∩ \{1,…j\}$ and hence the same set $P$ and the same value $D$.
    The number of row additions of \sgauss is bounded by $D$ by a similar argument (see \cite[Lemma 3]{DW:One-Block-per-Row:2019}). The analysis in \cite{DW:One-Block-per-Row:2019} proceeds by bounding $D$ in expectation and hence carries over to our case.\qedhere
\end{enumerate}
\end{proof}

\myparagraph{Efficiency.} While \sgauss and \ribbon are tied in $\O$-notation, \ribbon improves upon \sgauss in constant factors for the following reasons:
\begin{itemize}
    • There is no need to pre-sort the keys by $s(x)$.
    • \sgauss requires explicitly storing a pivot position for each row. %
    This is because \sgauss does not compute an echelon form but only ensures that in each row the left-most $1$-entry — the pivot — is the bottom-most $1$-entry of its column.
    • \sgauss performs roughly $D$ elimination steps that, depending on some bit, turn out to be xor-operations or no-ops. Ribbon on the other hand performs roughly $D/2$ bit shifts and $D/2$ (unconditional) xor operations. Though the details are complicated, intuition on branching complexity seems to favour ribbon.
\end{itemize}

\subsection{Ribbon with Smash}
\label{sec:smash}
When aiming for high space efficiency, there is an issue with early and late columns of the linear system. We shall describe the problem and its solution in an extreme but simple case where perfect space efficiency, i.e.\ $m = n$ is desired.

\begin{figure}[htb]
    \centering
    \begin{tabular}{c@{}cc@{}c}
         \textbf{(a)}&&\textbf{(b)}&\\[-5pt]
         & \includegraphics[page=2]{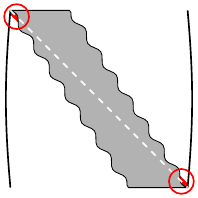} && \includegraphics[page=3]{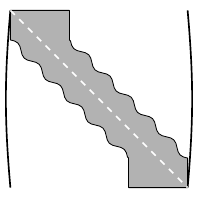}
    \end{tabular}
    \caption[fragile]{
        \textbf{(a)} Consider the diagonal (dashed) in a square Ribbon system. Its beginning and end may lie outside of the of the (shaded) ribbon area.\\ %
        \textbf{(b)} Our “smash” variant solves this problem.
    }
    \label{fig:ribbon-diagonal}
\end{figure}

In the absence of redundant equations, the construction of the linear system succeeds only if all slots of $M$ can be filled. For the first $i$ slots to be filled, it is necessary that $|\{x ∈ S \mid s(x) ≤ i\}| ≥ i$. \Cref{fig:ribbon-diagonal} \textbf{(a)} illustrates that random fluctuations make this unlikely. There is a similar problem relating to the last $i$ columns\footnote{To see the symmetry, we could have argued about the rank of the first $i$ columns of $M$ which is \emph{at most} $|\{x ∈ S \mid s(x) ≤ i\}|$.}. %

We can address this issue by artificially inflating the probabilities $\Pr[s(x) = 1]$ and $\Pr[s(x) = m-w+1]$ of the first and last starting position by a factor of $ℓ$ we call the \emph{smash value}. Such a distribution for $s$ is easy to implement using a uniform distribution on $[-ℓ+2,m-w+ℓ]$ and “clamping” the sampled value to $[1,m-w+1]$ using min and max functions. Micro-benchmarks show roughly 3ns overhead per query for smash, on an Intel Skylake CPU.

For a smash value of $ℓ = w/2$ and a ribbon width of $w = ω(\sqrt{n})$ the matrix diagonal is firmly within the ribbon, see \cref{fig:ribbon-diagonal} \textbf{(b)}.
It is not hard to prove that such a matrix is asymptotically as likely to be regular as a fully random $n × n$ matrix.
That probability is $c₂ ≈ 0.289$; see \cite{Cooper:Rank-Of-Random-Matrices:2000}.

In \cref{sec:standard-scalability} we present empirical findings showing that non-zero smash values also benefit success probabilities in practically more relevant cases with $ε > 0$ and $w = \smash{\O(\frac{\log n}{ε})}$. %

%% file: homog.tex
Recall that the idea underlying Ribbon filters is to pick hash functions $\svec{h} : \U → \{0,1\}^m$, $b : \U → \{0,1\}^r$ and find $Z ∈ \{0,1\}^{m×r}$ such that all $x ∈ S$ satisfy $\svec{h}(x)·Z = b(x)$, while most $x ∈ \U \setminus S$ will not.

We now examine what happens when we get rid of the fingerprint function $b$ effectively setting $b(x) = 0$ for all $x ∈ \U$. A filter is then given by a solution $Z$ to the \emph{homogeneous} system $(\svec{h}(x)·Z = 0^r)_{x ∈ S}$. The FP rate for $Z$ is $f_Z = \Pr_{a \sim H}[a·Z = 0^r]$ where $H$ is the distribution of $\svec{h}(x)$ for $x ∈ \U$. An immediate issue with the idea is that $Z = 0^{m × r}$ is a solution giving $f_Z = 1$. A solution $Z$ chosen \emph{uniformly at random} from all solutions fares better, however. To obtain one, all free variables, i.e.\ the variables corresponding to empty rows of $M$, are initialized randomly during back substitution.%
\footnote{Our implementation uses trivial pseudo-random assignments: a free variable in row $i$ is assigned $pi \bmod 2^{r}$ for some fixed large odd number $p$.}
The overall FP rate is then $f = \E[f_Z]$ where $Z$ depends on the randomness in $(\svec{h}(x))_{x ∈ S}$ and the free variables.

We call the resulting construction \emph{Homogeneous \ribbon filter}. It has two obvious advantages over Standard \ribbon filters:
\begin{itemize}
    • Constructions can never fail, regardless of $n$, $ε$ and $w$. This is simply because a homogeneous linear system always has at least the trivial solution. %
    • The absence of fingerprints slightly improves time and space in construction. (In optimized implementations, query times are essentially the same.)
\end{itemize}
A complication is that the FP rate $f$ can be higher than $2^{-r}$.
Intuitively, if too many equations constrain some part of $Z$ then that part will be insufficiently random.
For $εw > Cr$ (for some constant $C > 0$) and large $n$, this effect is negligible as we shall argue in \cref{thm:homogeneous-ribbon-space-efficient}. This still leaves us with two disadvantages, especially for small $n$ and high $r$:

\begin{itemize}
    • The product $εw$ must be proportional to $r$ (for acceptable $f$) whereas for Standard Ribbon $εw$ need only be proportional to $\log n$ (for acceptable success probability). We should therefore not expect an improvement over Standard Ribbon for $r = Ω(\log n)$.
    • For small $n$, the variance of $f_Z$ is quite high, meaning small filters will occasionally have significantly more false positives (e.g. due to random skew in $(s(x))_{x∈S}$) with no obvious way to detect this during construction. This could be dangerous for some applications.
\end{itemize}

\subsection{Analysis}

We can make a strong case for Homogeneous \ribbon filters by showing that arbitrarily small space overhead at arbitrarily large size $n$ is achievable. This neither requires the ribbon width $w$ to scale with $n$, nor a deviation from the pure construction (e.g.\ by partitioning the key set into small shards).

\def\SPACE{\textsc{space}}
\def\OPT{\textsc{opt}}
By space overhead we mean $\frac{\SPACE}{\OPT}-1$ where $\SPACE$ is the space usage of the filter in bits per key and $\OPT = -\log₂(\fpr)$ is the information-theoretic lower bound for filters that achieve the same FP rate.

\begin{theorem}
    \label{thm:homogeneous-ribbon-space-efficient}
  There exists $C ∈ ℝ^+$ such that for any $ε < 1/2$, any desired FP rate $ϕ > 0$ and $r$ the closest integer to $-\log₂(φ)$ the following holds.
  For any $w ∈ ℕ$ with $εw > C \max(\log w, r)$ and $n ∈ ℕ$ the Homogeneous \ribbon filter with $n$ keys and parameters $ε,w,r$ has $\fpr ∈ [ϕ/2,2ϕ]$ and space overhead at most $2ε$.
\end{theorem}

Our argument starts with the following simple observation.

\begin{lemma}
    \label{lem:homogeneous-fpr}
    In the context of a Homogeneous \ribbon filters let 
    $p$ be the probability that for $y ∈ \U \setminus S$ the vector $\svec{h}(y)$ is in the span of $(\svec{h}(x))_{x ∈ S}$. Then we have
    \[\fpr = p + (1-p)2^{-r}.\]
\end{lemma}

\begin{proof}
    First assume there exists $S' ⊆ S$ with $\svec{h}(y) = \sum_{x ∈ S'} \svec{h}(x)$ which happens with probability $p$. In that case \[\svec{h}(y)·Z = (\sum_{x ∈ S'}\svec{h}(x))·Z = \sum_{x ∈ S'}(\svec{h}(x)·Z) = 0\] and $y$ is a false positive. Otherwise, i.e.\ with probability $1-p$, an attempt to add $\svec{h}(y)·Z = 0$ to $M$ after all equations for $S$ were added would have resulted in a (non-redundant) insertion in some row $i$. During back substitution, only one choice for the $i$-th row of $Z$ satisfies $\svec{h}(y)·Z = 0$. Since the $i$-th row was initialized randomly we have $\Pr[\svec{h}(y)·Z = 0 \mid \svec{h}(y) ∉ \mathrm{span}((\svec{h}(x))_{x ∈ S)}] = 2^{-r}$.
\end{proof}

We shall now derive an asymptotic bound on $p$ in terms of large $w$ and small $ε$ (recall $ε = \frac{m-n}{n}$). It is too imprecise to estimate $p$ and $\fpr$ in practical settings, which we do empirically in \cref{sec:homog-in-practice}. The main takeaway is rather that $p$ does not depend on $n$. We may therefore expect Homogeneous \ribbon filters to scale to arbitrary sizes $n$ with no increase in $\fpr$ even when $w$ and $ε$ are constants.

\begin{lemma}
    \label{lem:extra-fpr-homogeneous}
    There exists a constant $C$ such that for any $w ∈ ℕ$ and $C\frac{\log w}{w} ≤ ε ≤ \frac 12$ we have $p = \exp(-Ω(εw))$.
\end{lemma}

The main ingredient in the proof is that $\exp(-Ω(εw))$ bounds the number of keys that cannot be (non-redundantly) inserted, which follows from \cite{DW:One-Block-per-Row:2019}.\footnote{Recall \cref{fn:no-leading-1-in-theory}.}

\begin{proof}[Proof Sketch.]
\def\pos{\mathrm{pos}}
    We may imagine that $S ⊆ \U$ and $y ∈ \U \setminus S$ are obtained from a set $S^+ ⊆ \U$ of size $n+1$ by picking $y ∈ S^+$ at random and setting $S = S^+ \setminus \{y\}$. Then $p$ is simply the expected fraction of keys in $S^+$ that are contained in some \emph{dependent set}, i.e. in some $S' ⊆ S^+$ with $\sum_{x ∈ S'} \svec{h}(x) = 0^m$. Clearly, $x$ is contained in a dependent set if and only if it is contained in a \emph{minimal} dependent set.
    Such a set $S'$ always touches a consecutive set of positions, i.e.\ $\pos(S') := \bigcup_{x ∈ S'} [s(x),s(x) + w -1 ]$ is an interval.
    
    We call an interval $I ⊆ [m]$ \emph{long} if $|I| ≥ w²$ and \emph{short} otherwise. We call it \emph{overloaded} if $S_I := \{x ∈ S^+ \mid s(x) ∈ I\}$ has size $|S_I| ≥ |I|·(1-ε/2)$.
    Finally, we call a position $i ∈ [m]$ \emph{bad} if one of the following is the case:
    \begin{enumerate}[(\textsc{b}1)]
        •\label{it:b1} $i$ is contained in a long overloaded interval.
        •\label{it:b2} $i ∈ \pos(S')$ for a minimal dependent set $S'$ with long non-overloaded interval $\pos(S')$.
        •\label{it:b3} $i ∈ \pos(S')$ for a minimal dependent set $S'$ with short interval $\pos(S')$.
    \end{enumerate}
    We shall now establish the following
    \[\textbf{Claim: } ∀i ∈ [m]: \Pr[i \text{ is bad}] = \exp(-Ω(εw)).\]
    For each $i ∈ [m]$ the contributions from each of the badness conditions (\textsc{b1,b2,b3}) can be bounded separately. In all cases we use our assumption $ε ≥ C\smash{\frac{\log w}{w}}$. It ensures that $\exp(-Ω(εw))$ is at most $\exp(-Ω(\log w)) = \smash{w^{-Ω(1)}}$ and can “absorb” factors of $w$ in the sense that by adapting the constant hidden in $Ω$ we have $w \exp(-Ω(εw)) = \exp(-Ω(εw))$.
    
    \makeatletter
    \def\smallunderbrace#1{\mathop{\vtop{\m@th\ialign{##\crcr
       $\hfil\displaystyle{#1}\hfil$\crcr
       \noalign{\kern3\p@\nointerlineskip}%
       \tiny\upbracefill\crcr\noalign{\kern3\p@}}}}\limits}
    \makeatother
    \begin{enumerate}[(\textsc{b}1)]
        • A Chernoff bound for sums $X = \sum_{j} X_j$ of i.i.d.\ indicator random variables with $μ = \E[X]$ is
        \begin{equation}
            \Pr[X ≥ (1+δ)μ] ≤ \exp(-δ²μ/3). \label{eq:chernoff}
        \end{equation}
        We use it in a case where $I$ is an interval and $X₁,…,X_{n+1}$ indicate which of the keys in $S^+$ have a starting position within $I$. For $n \gg w$ we have
        \[ μ = \E[X] ≤ \frac{(n+1)|I|}{m-w+1} ≈ \frac{n|I|}{m} = |I|/(1+ε). \]
        Skipping over some uninteresting details, the probability for $I$ to be overloaded is (for $n \gg w$)
        \begin{align}
            \Pr[X &≥ (1-ε/2)|I|] ≤ \smash{\Pr[X ≥ (1+\smallunderbrace{ε/6}_{δ})\smallunderbrace{|I|/(1+ε)}_{≥ μ}]}\notag\\
            \refrel{eq:chernoff}{Eq.}{≤}
            &\exp(\frac{-ε²|I|}{108(1-ε)}). %
            \label{eq:chernoff-long-overloaded}
        \end{align}
        The probability for $i ∈ [m]$ to be contained in a long overloaded interval is bounded by the sum of \cref{eq:chernoff-long-overloaded} over all lengths $|I| ≥ w²$ and all $|I|$ offsets that $I$ can have relative to $i$.
        The result is of order $\exp(-Ω(ε²w²)$ and hence small enough.
        • Consider a long interval $I$ that is not overloaded, i.e.\ $|I| ≥ w²$ and $|S_I| ≤ (1-ε/2)|I|$. There are at most $2^{|S_I|}$ sets $S'$ of keys with $\pos(S') = I$ and each is a dependent set with probability $2^{-|I|}$ because each of the $|I|$ positions of $I$ that $S'$ touches imposes one parity condition.
        
        The probability for $I$ to support at least one dependent set is therefore at most $2^{-|I|}·2^{|S_I|} = 2^{-\frac{ε}{2}|I|} = \exp(-Ω(ε|I|))$.
        
        Similar as in (\textsc{b1}) for $i ∈ [m]$ we can sum this probability over all admissible lengths $|I| ≥ w²$ and all offsets that $i$ can have in $I$ to bound the probability that $i$ is bad due to (\textsc{b2}).
        •
        \def\Sred{S_{\mathrm{red}}}
        Let $\Sred ⊆ S$ be the set of \emph{redundant} keys, i.e.\ keys for which \cref{algo:ribbon-insertion} returns “\textsc{success} (redundant)”. While $\Sred$ depends on the insertion order, the \emph{rank defect} $|\Sred| = n - \mathrm{rank}((\svec{h}(x))_{x ∈ S})$ does not.
        A central step in \cite{DW:One-Block-per-Row:2019} implies that $\E[|\Sred|] = m·\exp(-Ω(εw))$.\footnote{%
        The bound is only used to show that for $w = Ω(\frac{\log n}{ε})$ all insertions succeed with high probability.}
        
        Now if $i$ is bad due to (\textsc{b3}) then $i ∈ \pos(S')$ for some minimal dependent set $S'$ with short $\pos(S')$. At least one key from $S'$ is redundant (even if all keys from $\pos(S')$ are inserted first). In particular, $i$ is within short distance ($< w²$) of the starting position of a redundant key $x$. Therefore at most $|\Sred|·2w²$ positions are bad due to (\textsc{b3}), which is an  $\exp(-Ω(εw))$-fraction of all positions as desired.
    \end{enumerate}
    With a concentration argument the following variant of the claim can be proved. We omit the details.
    \[\textbf{Claim': } ∀x ∈ S^+: \Pr[s(x) \text{ is bad}] = \exp(-Ω(εw)).\]
    Now assume that a key $x∈ S^+$ is contained in a minimal dependent set $S'$.
    It follows that all of $\pos(S')$ is bad. Indeed, either $\pos(S')$ is a short interval ($→$ \textsc{b3}) or it is long. If it is long, then it is overloaded ($→$ \textsc{b1}) or not overloaded ($→$ \textsc{b2}). In any case $s(x) ∈ \pos(S')$ is bad.
    
    Therefore, the probability $p$ for $y ∈ S^+$ to be contained in a dependent set is at most the probability for $s(y)$ to be bad. This is upper-bounded by $\exp(-Ω(εw))$ using Claim'.
\end{proof}

We are now ready to prove \cref{thm:homogeneous-ribbon-space-efficient}.
\begin{proof}[Proof of \cref{thm:homogeneous-ribbon-space-efficient}]
  We shall find that our filter has $\fpr ≤ 2^{-r}(1+ε²)$ and hence $\fpr ∈ [ϕ/2,2ϕ]$ with high probability.
  
  The space consumption of $Z ∈ \{0,1\}^{m × r}$ is $\SPACE = \frac{mr}{n} = r(1+ε)$ bits per key. To relate this to $\OPT$, we need bounds on $p$.
  
  By assumption $εw > C \log w$, so a large enough choice for $C$ permits the use of \cref{lem:extra-fpr-homogeneous}, which guarantees $p = \exp(-Ω(εw))$.
  Again using $εw > C\max(\log w, r)$ for large enough $C$ gives
  \begin{equation}
    \label{eq:upper-bound-on-p}
    p ≤ \exp(-2\log(w)-r) ≤ \frac{1}{w²}e^{-r} ≤ ε²2^{-r}.
  \end{equation}
  Together with \cref{lem:homogeneous-fpr} we get
  \begin{align*}
    \OPT &= -\log₂(\fpr)
    \refrel{lem:homogeneous-fpr}{Lem}{=}
    -\log₂(p + (1-p)2^{-r})\\
    &≥ -\log₂(p+2^{-r})
    \refrel{eq:upper-bound-on-p}{Eq.}{≥}
    -\log₂(2^{-r}(1+ε²))\\
    &=r - \log₂(1+ε²) ≥ r-ε².
  \end{align*}
  Putting everything together yields
  \[
    \frac{\SPACE}{\OPT} = \frac{r(1+ε)}{r-ε²} ≤ \frac{(1+ε)}{(1-ε²)} ≤ 1+2ε.
  \]
  The last step makes use of $ε < \frac{1}{2}$.
\end{proof}

\begin{figure}
    \centering
    \includegraphics[page=4]{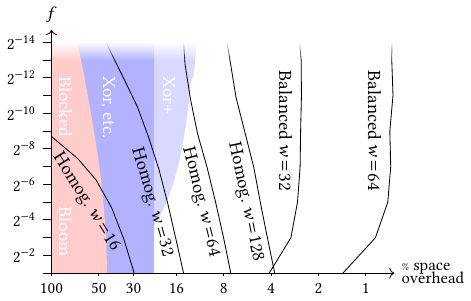}
    \caption{Combinations of space overhead and false positive rate achievable with various ribbon widths $w$ and large $n$. (Experimental data for Ribbon configurations use $n=10^{6}$ but generalize.) }
    \label{fig:reach-of-ribbon-configs}
\end{figure}

\subsection{Space efficiency in practice}
\label{sec:homog-in-practice}
To minimize the space overhead $r(1+ε)/\log\fpr^{-1} - 1$ for chosen values of $r$ and $w$ in a Homogeneous Ribbon filter, $ε$ must be neither too large or too small. (Small $ε$ causes $f$ to explode due to densely-packed constraints.)
To choose $ε$, we turn to simulations on random data, building large structures and testing the FP rate. Using $m = 3·10^{7}$ (among others), $w ∈ \{16,32,64,128\}$, and $r ∈ [1,16]$, we see a pretty clear pattern in where the space overhead is minimized for any large $n$:
\begin{equation}
\label{eq:homog-recommended-epsilon}
    ε \approx \frac{4 + r/4}{w}.
\end{equation}
Note how the recommendation $ε > C \max(\log w, r)/w$ one might derive from \cref{thm:homogeneous-ribbon-space-efficient} (vaguely) agrees.
We use (\ref{eq:homog-recommended-epsilon}) in all experiments. In \cref{fig:reach-of-ribbon-configs} we show combinations of space overhead and FP rate to expect from Homogeneous Ribbon filters for $w ∈ \{16,32,64,128\}$ and large $n$.\footnotemark\ 
Balanced Ribbon filters, also in \cref{fig:reach-of-ribbon-configs}, are discussed in \cref{sec:balanced}.

For example, consider using $r = 7$ for roughly 1\% FP rate and $w=64$ for reasonable space-vs.-time trade-off. Using $ε \approx 0.09$ from \cref{eq:homog-recommended-epsilon} we observe $\fpr \approx 0.81\% > 0.78\% \approx 2^{-7}$, so actual space overhead is closer to 10\% than the 9\% allocated with $ε$.

\footnotetext{Standard Ribbon filters are not shown because the achievable overhead depends on $n$. We do recommend Standard Ribbon (with smash) for small $n < 10^4$ where Homogeneous Ribbon filters have high variance in FP rate.}

%% file: configurability.tex
\subsection{Configurability and elasticity}
\label{sec:arbitrary}
A useful and perhaps previously unreplicated feature of Bloom filters is the ability to efficiently utilize any amount of space for minimizing the FP rate in representing any number of keys. We call this \emph{configurability} and suggest it is practically important for space efficiency. Consider an application with little control over the number of keys going into a filter. Even if we use a perfectly space efficient filter for a specific FP rate, we could be wasting significant space due to internal fragmentation from an allocator. A memory allocator like jemalloc~\cite{JemallocManual} averages about 10\% internal fragmentation on arbitrarily sized allocations, space that should ideally be used by the filter to reduce its FP rate~\footnote{See RocksDB's \texttt{optimize\_filters\_for\_memory} option~\cite{RocksDBWikiBloom}.}.

More specifically, Bloom alternatives such as Cuckoo, Quotient, and Xor conventionally use cells of some whole number of bits, as Ribbon does with $r$. Whole number $r$ limits space-efficient choices of FP rates and bits per key. For example, using only $5$-bit cells when $5.5$ bits per cell is available adds roughly 10\% space overhead to our filter. %
Other than Ribbon, which is tied to the two-element field, these same Bloom alternatives can use fractional-bit cell sizes. Some configurations can even be made efficient, such as $64/i$ bits per cell for whole $i$ (an existing Xor10.666 implementation is tested in \cref{sec:experiments}), but fine granularity would surely be more CPU intensive.

An alternative way of generalizing to \emph{effectively} fractional $r$ is to split available space between two structures: one using $⌈r⌉$ solution columns (or bits per cell) and the other using $⌊r⌋$ for a weighted average of $r$, chosen to fit available space. This only slightly increases the overall space overhead. For example, using $r=5.5$ yields (non-homogeneous) FP rate of $3/128$. The lower bound for this rate is $λ = 5.415$ bits per key, so the approach adds 1.57\% to overall space overhead. This addition is a larger 6.00\% for $r=1.5$, or smaller 0.82\% for $r=10.5$ bits per key. Practical concerns with splitting into two structures includes (a) essentially doubling many of the space usage penalties associated with small structures (when applicable), and (b) independently seeding hashes or accepting joint construction success probability (when applicable).

For Ribbon we recommend a variant of that approach within a single structure: using only $⌊r⌋$ solution bits per row for some prefix of rows, and $⌈r⌉$ bits per row for the rest of rows. The banding process (\cref{algo:ribbon-insertion}) is unchanged, but small changes are needed to back-substitution and query (more details in~\cref{sec:solution-layout}). Because of Ribbon's locality of probes in queries, unlike standard Xor filters, a diminishingly small $w/m$ portion of queries cross the boundary between $⌊r⌋$ and $⌈r⌉$ columns (utilizing only $⌊r⌋$ in such cases), so space efficiency is very close to the idealized split approach, and probably better in practice: around 1\% additional space overhead for common configurations (see e.g. $r=7.7$ in \cref{sec:experiments}).

Although the split approach enables near-continuous configurability for many kinds of filters, the single-structure approach for Ribbon filters has an advantage we call \emph{elasticity}, for applications like ElasticBF~\cite{LTGLX:ElasticBF:2019}. Like an Xor filter, one can drop entire columns from a Ribbon filter, for a corresponding higher FP rate. With Ribbon filters, we also have the ability to drop part of the last column, so a finished filter can be trimmed down with bit granularity. Generalizing further, a finished Ribbon filter can be split to several smaller structures with independent FP rates\footnote{For Homogeneous Ribbon, the portion of the FP rate from degradation is not independent.}, with product as small as the FP rate of the starting structure. Similarly, a finished Ribbon filter could be physically (re-)partitioned at arbitrary boundaries by duplicating as little as $(w - 1)r$ bits at each partition boundary such that each query accesses only one partition.

%% file: solution-layout.tex
\subsection{Solution structure layout}
\label{sec:solution-layout}
Here we examine memory layouts for the solution matrix $Z ∈ \{0,1\}^{m × r}$, which is critical for fast Ribbon filter queries. Prior work~\cite{DW:One-Block-per-Row:2019} only evaluated the $r=1$ case, where $Z$ is a Boolean (bit) vector.

Xor filters conventionally use \textbf{row-major layout} of the solution structure, wherein the whole row $i$ of $Z$ immediately precedes the whole row $i+1$ in memory. A $w=64$ Ribbon filter combines roughly an order of magnitude more rows, conditionally, than a standard Xor filter combines unconditionally (three rows). In fact, with number of solution columns commonly $5 \leq r \leq 15$, standard Xor filters typically access more columns than rows, while Ribbon filters typically access more rows than columns. Although row-major layout could likely be made efficient for Ribbon in some special cases using SIMD, we do not find it generally workable for a fast and highly-configurable filter.

The opposite is \textbf{column-major layout}, in which the entire column $i$ of $Z$ precedes column $i+1$ in memory. Column-major is essentially ideal for querying a single result bit for a key, as we only have to access a ``contiguous'' (usually unaligned) $w$ bits, bitwise-AND with $c(x)$, and get the bit parity. The problem with column-major is that accessing more result bits is not an adjacent memory access. Although the several memory addresses are easily computed and can be fetched in parallel, some testing shows this to be be relatively expensive for $r > 2$ for a filter that is not hot in memory cache.

\begin{figure}
    \centering
    \includegraphics[page=7]{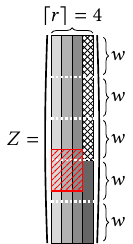}\hspace{1cm}
    \includegraphics[page=8]{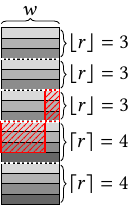}
    \caption[A solution matrix (left) and $w$-bit interleaved column-major layout (ICML, right) of that matrix, mixing $⌊r⌋$ and $⌈r⌉$ columns as in \cref{sec:arbitrary}.]{
        A solution matrix (left) and $w$-bit interleaved column-major layout (ICML, right) of that matrix, mixing $⌊r⌋$ and $⌈r⌉$ columns as in \cref{sec:arbitrary}. The \tikz{\draw[draw=red,pattern color=red,pattern=south west lines,thick] rectangle(3ex,1.5ex);} shaded region shows the bits used in a query crossing the boundary between $⌊r⌋$ and $⌈r⌉$.
    }
    \label{fig:icml}
\end{figure}

Our preferred solution layout for Ribbon filters is \textbf{interleaved column-major layout} (ICML), because it has locality very close to row-major and decoding efficiency very close to column-major. The memory space is divided into conveniently sized ICML words, and grouped into blocks of $r$ words. Each block is the column-major layout of some contiguous rows of $Z$. See \cref{fig:icml}, which generalizes this layout to a mixed number of columns for fractional $r$.

For ICML word size equal to $w$, at most and almost always two words are combined for reconstructing each result bit. This means that the amount of adjacent memory accessed for a full query is $2rw$ bits, while the ideal minimum is $rw$ bits. For example, with $r=6$ and $w=64$, ICML accesses 768 bits per query, with 384-bit alignment, which translates to an average of accessing $2.25$ Intel cache lines (512 bits) and essentially 1 page (4KB) per query; with only 6-bit ``alignment,'' row-major would access $1.5$ cache lines per query. A standard Xor filter accesses essentially 3 cache lines and nearly as many pages per query.

Ribbon back-substitution is an especially fast, streaming operation for layouts based on column major. We can buffer $w$ rows of $Z$ in $r$ temporary values of width $w$, likely fitting in CPU registers, and use those buffers for (a) computing the logical previous bit for each column, and (b) flushing to our solution structure for each $w$ rows ($rw$ bits).

As is well known for Bloom filters, queries can potentially be optimized with short-circuiting: returning from a ``negative'' query as soon as a probed bit is zero, ensuring the query must return \false. A similar approach works for Ribbon filters using layouts based on column-major, returning as soon as a result bit does not match expectation. Although cache-local Bloom filters are so optimized that this approach rarely pays off any longer~\cite{LNKB:Bloom:2019}, our Ribbon implementation uses short-circuiting except for compile-time fixed $r ≤ 4$. The distinction is visible in observed query time ranges in \cref{sec:experiments}.

We also like the clean configurability of layouts based on column-major. Parameter $r$ should be freely chosen to balance FP rate vs. space usage, and that choice is much more free when it does not affect instruction-level data alignment, only alignment in CPU caches and pages, which we consider a relatively minor concern. Ribbon-width ICML is good for using $⌊r⌋$ columns before $⌈r⌉$ columns (\cref{sec:arbitrary}), because upon determining the starting memory location and (smallest) applicable number of columns, which can be done without conditional branches, the remaining query code does not have to be aware of mixed numbers of columns; see~\cref{fig:icml}. ($⌈r⌉$ before $⌊r⌋$ is better for pure column major.) 

A minor disadvantage of ICML is that the number $m$ of solution rows must be a multiple of the number of bits in an ICML word, which can present a conflict between configurability (accommodating any number of keys) and space efficiency for small $n$.%

%% file: practical-hashing.tex
\subsection{Practical hashing for Ribbon}
\myparagraph{Hash expansion.} A filter structure ``consuming'' some quantity of hash information can operate from a smaller hash $\mathcal{H}(x)$ of the original key $x$~\cite{DM:Bitstate:2004,DM:Bloom:2004}. The practical requirements for Ribbon filters are these: %
\begin{itemize}
    \item $\mathcal{H}(x)$ values must be large enough to have an insignificant baseline FP rate due to full hash collisions, i.e.\ $\mathcal{H}(x_1) = \mathcal{H}(x_2)$ for $x_1 ≠ x_2$. A 64-bit hash for $\mathcal{H}$ should suffice for almost all non-cryptographic applications, as having $2^{32}$ keys in a single filter incurs a baseline FP rate of just $2^{-32}$.
    \item $\mathcal{H}(x)$ is effectively extended / expanded / remixed to what is consumed\footnote{In at least two cases~\cite{Dillinger:RocksDBIssue4120,A:Partitioned:2020}, implementations citing an asymptotic result for efficient hashing in Bloom filters~\cite{kirsch2008less} had practical flaws that previous work \cite{DM:Bloom:2004} warned about.
    }. For Ribbon, it is most important to minimize correlations between the starting location and other hash consumers. A starting location computed with $\fastrange$~\cite{L:Fastrange:2019} on $\mathcal{H}(x)$ relies primarily on upper bits, so multiplying $\mathcal{H}(x)$ by a large odd constant (as with Knuth multiplicative hash \cite{K:Sorting:1998}) seems to suffice for removing correlation. Details are in the reference implementation of Ribbon~\cite{Dillinger:RocksDBRibbon}.
\end{itemize}

\myparagraph{Re-seeding.} 
Some Ribbon designs need the ability to retry construction with sufficiently independent hashing to have an independent probability of construction success. Observe that for Ribbon filters a full hash collision does not interfere with construction success (it only produces a redundant equation). We find in significant testing that modifying an unseeded stock hash value with simple XOR with a pseudorandom seed then multiplication by a large odd constant suffices for independent probability of construction success. See \cite{Dillinger:RocksDBRibbon} for details. Assuming uniform hashes, an effective alternative to re-seeding on failed construction is simply to increase $m$ by a factor of $\frac{w+1}{w}$.

%% file: standard-scalability.tex
\subsection{Standard Ribbon scalability}
\label{sec:standard-scalability}
We refer to the non-homogeneous Ribbon construction, including \emph{smash} when appropriate, as \emph{Standard Ribbon}. Construction fails with some probability depending on $m$, $n$ and $w$, though we have no formula. \cref{tab:overheads} provides some empirical data points for how much configured space overhead, $ε = (m-n)/n$, is required for several failure probabilities that represent different construction time vs. solution space trade-offs. Observe that unlike standard Xor filters, Ribbon does not exhibit sharp threshold behavior in construction success; almost sure construction success with Standard Ribbon requires significantly more space overhead than 5\% failure chance, a good space-time trade-off in our judgment.

\def\mean{$\langle$add till failure$\rangle$}
\begin{table*}[ht]
 \caption{Standard Ribbon space overhead $ε$ from empirical data}
 \label{tab:overheads}
 \begin{tabular}{lc|ccccccl}
  \toprule
  Ribbon & Failure     & $w/2$ \emph{smash} & $w/4$ \emph{smash} & 0 \emph{smash} &  &  &  & Each additional \\
  width  & probability & $m=2^{10}$ & $m=2^{10}$ & $m=2^{10}$ & $m=2^{14}$ & $m=2^{17}$ & $m=2^{24}$ & doubling of $m$ \\
  \midrule
  $w=128$ & 0.5        & 0.2\%       & 0.1\%       & 1.0\%     & 1.1\%      & 2.2\%      & 4.7\% & +0.38\%\\
  $w=128$ & \mean      & 0.2\%       & 0.2\%       & 1.1\%     & 1.2\%      & 2.3\%      & 4.8\% & +0.38\% \\
  $w=128$ & 0.05       & 0.5\%       & 0.5\%       & 2.2\%     & 2.6\%      & 3.7\%      & 5.9\% & +0.38\% \\
  $w=128$ & 0.001      & 1.1\%       & 1.2\%       & 4.1\%     & 4.6\%      & 5.8\%      & > 7\% & +0.38\% \\
  \midrule
  $w=64$ & 0.5         & 0.3\%       & 0.4\%       & 2.0\%     & 3.8\%      & 6.3\%      & 11.7\% & +0.83\% \\
  $w=64$ & \mean       & 0.8\%       & 0.8\%       & 2.2\%     & 4.1\%      & 6.5\%      & 12.1\% & +0.83\% \\
  $w=64$ & 0.05        & 3.7\%       & 2.9\%       & 4.8\%     & 7.0\%      & 9.4\%      & 15.0\% & +0.83\% \\
  $w=64$ & 0.001       & 11.4\%       & 7.1\%       & 9.2\%     & 11.5\%      & 13.8\%      & > 19\% & +0.83\% \\
  \midrule
  $w=32$ & \mean       & 5.9\%        & 5.2\%       & 6.3\%     & 13.2\%      & 19.2\%      & 35.2\% & +2\% \\
  \midrule
  $w=16$ & \mean       & 28.9\%       & 27.0\%      & 27.5\%    & 67.3\%      & > 100\%      & $\gg$ 100\% & +??\% \\
  \bottomrule
 \end{tabular}
\end{table*}

Although Standard Ribbon does not scale infinitely for fixed ribbon width and space overhead, it is more space-efficient than standard Xor (23\% overhead) for most practical $n$, which can be seen in \cref{tab:overheads} and \cref{sec:experiments}. Unlike many other Gaussian structures, construction speed is not a significant concern for scaling Ribbon to large $n$.

\myparagraph{Scaling with sharding.} There are many standard or obvious ways to construct a large, space-efficient filter from many smaller space-efficient filters~\cite{Putze:Efficient-Bloom-Filters:2009,Vigna:Fast-Scalable-Construction-of-Functions:2016}. Two ways of leveraging Ribbon features are notable, but not evaluated in detail:
\begin{itemize}
    \item If uniformly hash-partitioning keys into fixed-size data structure shards, the \emph{fractional $r$} feature of Ribbon (\cref{sec:arbitrary}) can be used to accommodate variance in the number of keys mapped to each shard.
    \item If determining hash ranges to assign to each shard, on-the-fly banding allows adding entries or buckets (in sharding hash order, independent of start location order) until one fails and starts the next shard. The ``\mean'' rows in~\cref{tab:overheads} correspond to this strategy of adding entries until one fails, so yields good average space efficiency without construction retries, such as $< 1\%$ overhead with $m = 2^{10}$ per shard, not including sharding metadata.
\end{itemize}

\myparagraph{Shard sizes.} \cref{sec:smash} describes an inherent unlikelihood of filling all slots in a Ribbon system, even if $m = w$, and how the likelihood is similar with $m = \Theta(w^{2})$. Because
the expected number of empty slots at first failure to add remains constant even for small $m$,
the median proportion of unoccupied slots at failure decreases with $m$ before increasing with $m$, for a fixed ribbon width $w$. For common ribbon widths, the minimum appears to be around $w^{2}/4$, which we suggest is the \emph{natural} shard size, subject to practical adjustment for the application.

\myparagraph{Soft sharding.} With Ribbon we can apply sharding at a higher abstraction layer than memory space, for potentially better space efficiency. In a typical \emph{hard sharding}, construction optimizes for each shard either (a) a set of keys, (b) a memory size, (c) a hash seed, or (d) some other configuration parameters, based on the others, and records the optimized configuration in some metadata. The change with \emph{soft sharding} is that each shard is assigned a contiguous range of Ribbon start locations (from a single Ribbon system) rather than a contiguous memory space (containing an independent Ribbon system). This should be a pure win for space efficiency, because the overlap of $w-1$ probing rows between adjacent soft shards allows them to, in effect, borrow some space from each other without expending metadata. (A Standard or Homogeneous Ribbon filter is a naive soft sharding with no metadata guiding the shard assignments.) With some ordering constraints and temporary tracking data, the Ribbon algorithm allows us to backtrack, such as for changing the hash seed within a shard or key assignments to shards. We do not analyze the ``soft sharding'' design space in detail, but use the idea for Balanced Ribbon.

%% file: balanced-ribbon.tex
\section{Balanced Ribbon}
\label{sec:balanced}

Balanced Ribbon is an experimental design for scaling and space-optimizing Standard Ribbon; for implementation details see~\cite{Dillinger:fastfilter:2021}. We intend Balanced Ribbon as an example in the design space opened up by the new on-the-fly and incremental Gaussian elimination algorithm, and encourage follow-up work to explore, optimize, and analyze this design space.

Balanced Ribbon extends Standard Ribbon with soft sharding and a new balanced allocation scheme tailored to this domain (related:~\cite{ABKU:Balanced:1999,CRS:Perfectly:2003,BFHM:WeightedBallsBins:2008,M:PowerOfTwo:2001,BKSS:BallsBinsOptimalLoad:2013,W:HashingLoadBalancing:2017}). Like many other hashing schemes, we start with the idea that each entry has two possible locations in the Ribbon, given by two hashes: an earlier primary location and a later secondary location in a distinct shard. The shard with the primary location is constructed before the shard with the secondary location and accommodates the key if possible. If not, we say the key is “bumped” and must be accommodated in its secondary shard, so shards add “bumped” entries first for best chance of success.
Metadata is constructed to indicate which keys are bumped\footnotemark.
The construction is greedy in that a shard tries to accommodate as many keys as possible, without considering where keys will be bumped to, and constructed shards are never revisited. Difficulty arises if using two uniform hashes, the smaller primary and the larger secondary: later shards are dominated by entries in their secondary location.

\footnotetext{There is no such disambiguation in \cite{DW:Retrieval-log-extra-bits:2019} and a query would combine information from \emph{both} locations. This significantly complicates the linear system, however.}

\myparagraph{Organizing shards for bumping.} To make this work we organize the soft shards into levels $1 .. \ell$, with level $i$ containing exactly $\lceil 2^{\ell - i - 1} \rceil$ shards, so we assume a power of two number of shards overall, $s = 2^{\ell - 1}$. Unlike some multi-level hashing schemes~\cite{BK:Multilevel:1990,KTC:Peacock:2008,KM:OneMoveHashing:2010}, an entry's primary location can be on any level $i$, with its secondary location uniformly on level $\min(i+1,\ell)$. Because no secondary locations are in level 1, we overload it with primary locations; level 1 shards have relative weight $1 + \alpha$ for primary locations and other level shards have weight $1 - \alpha$. See~\cref{fig:balanced}. With average $n/s$ keys per shard, we choose $\alpha \approx 3.5 / \sqrt{n/s}$ to ensure a sufficient supply of entries even for shards with three Poisson standard deviations below the mean number of entries. (We use $\alpha = 1/8$ for $n/s \approx 1000$.) Assuming we allocate our space overhead perfectly, the average number of entries bumped from each level 1 shard will be $\alpha n/s$. Because level 2 shards are half as many, they receive $2\alpha n/s$ entries on average for adding in secondary location. With those bumped entries, level 2 shards are now overloaded to relative weight $1 + \alpha$ compared to $1 - \alpha$ for later shards. With this, we have a recursive structure to ensure a continuous supply of entries eligible for bumping down to the last shard. (Like spatial coupling~\cite{W:SpatialCoupling:2021} and ``always go left''~\cite{V:How_Asymmetry:2003}, we are making productive use of \emph{less} randomness.)

\begin{figure}
    \centering
    \include{balanced}
    \caption[Bumping behavior between levels of Balanced Ribbon.]{
        Bumping behavior between levels of Balanced Ribbon. The expected relative quantity of entries with primary locations in each level is indicated by \tikz{\draw[pattern=vertical lines, pattern color=gray] rectangle(3ex,1.7ex);} where \tikz{\draw[pattern=south west lines, pattern color=red] rectangle(3ex,1.7ex);} are bumped and added in their secondary location \tikz{\draw[pattern=dots, pattern color=blue] rectangle(3ex,1.7ex);}. Not shown: allocation overheads ($ε$), Poisson variances, bucket boundaries, and dispersion within each shard.
    }
    \label{fig:balanced}
\end{figure}

The last shard (level $\ell$) is different but does not need to be complicated. If we configure our allocated space overhead assuming Standard Ribbon overheads for the last shard, along with tighter overheads for the other shards, it seems to work (single shard Balanced Ribbon $\equiv$ Standard Ribbon). For large number of shards, we observe the Balanced Ribbon final shards either completely overwhelmed with bumped entries (construction failure) or receiving almost no bumped entries. We believe this is because the overall variance in utilization of slots in all prior shards is concentrated into the last shards, and that variance is large relative to a single shard. Because the variance is small overall with a large number of shards, construction success is more predictable at scale (apparent threshold behavior).

\myparagraph{Buckets for bumping.} We use another (semi-)independent hash to order or group keys within a shard strictly for bumping. Selecting a threshold on that hash works well for both Ribbon and metadata space efficiency. First, we have chosen our shard size such that controlling the number of entries going into a shard is much more important than the particular set of entries. Second, a $\Theta(\log(n/s))$ bit threshold value is small metadata per shard. However, we do not want to incur the CPU time for sorting entries into so many buckets per shard.

Instead of a threshold we use $\Theta(\log(n/s))$ buckets of geometrically distributed sizes that can be selected independently, using the incremental feature of \ribbon to backtrack on failed buckets. Using some bit tricks to approximate a geometric distribution with $p=2^{-0.5}$ seems better than $p=2^{-1}$, perhaps due to variance in actual bucket sizes. We prefer 8 buckets per shard for 8 bits of metadata per shard, 0.008 bits per key for common shard size. A subtle part of maximizing space efficiency with independent buckets and soft sharding is to attempt adding a larger (in expectation) bucket in shard $i+1$ before attempting to add a smaller (in expectation) bucket in shard $i$, if the two shards are in the same level. Successfully adding the smaller could overflow enough to make it impossible to add the larger; on average, the better greedy choice is trying to add the larger bucket first. For CPU efficiency, we skip attempts to add a bucket that is very likely to fail given the number of successful additions to the shard.

\myparagraph{Overall.} Balanced Ribbon construction resembles an external sort between CPU cache and main memory. Queries depend on just one bit of sharding metadata, out of a typical 8 bits per shard. Space usage for $w=64$ Balanced Ribbon is $1.005 λ + 0.008$ bits per key, relative to the information-theoretic lower bound $λ$, as shown in \cref{fig:reach-of-ribbon-configs}. We have tested Balanced Ribbon with these space efficiencies up to 4 billion keys; arbitrary scaling might require natural increases in $w$ ($w \sim \log n$ and thus $n/s \sim \log²n$) or might benefit from design changes.

%% file: balanced.tex
{
\tikzstyle{shortenedArrow}=[<->,shorten >=0.5pt,shorten <=0.5pt]
\def\w{6}
\def\h{2}
\def\g{0.2}
\begin{tikzpicture}
\draw (0,0) rectangle ({\w / 2}, \h);
\draw ({\w / 2},0) rectangle ({\w * 3 / 4}, \h);
\draw ({\w * 3 / 4},0) rectangle ({\w * 7 / 8}, \h);
\draw ({\w * 7 / 8},0) rectangle (\w, {\h - \g*\h});

\node[left,yshift=-7pt] (blah1) at (0,0) {Level};
\draw[shortenedArrow,yshift=-7pt] (0,0) -- node[below] {1} ({\w / 2},0);
\node[left,yshift=-24pt] (blah1) at (0,0) {Shard};
\foreach \i in {1, 2, 3, 4, 5, 6, 7}
    \draw[shortenedArrow,yshift=-24pt] ({\w * (\i - 1) / 8},0) -- node[below] {\i} ({\w * \i / 8},0);
\draw[shortenedArrow,yshift=-24pt] ({\w * 7 / 8},0) -- node[below] {$s = 8$} (\w,0);

\fill[pattern=vertical lines, pattern color=gray] (0,0) rectangle ({\w / 2}, \h);

\fill[pattern=vertical lines, pattern color=gray] ({\w / 2},0) rectangle (\w, {\h - 2 * \g * \h});
\draw[shortenedArrow,yshift=-7pt] ({\w / 2},0) -- node[below] {2} ({\w * 3 / 4},0);
\fill[pattern=vertical lines, pattern color=gray] ({\w * 3 / 4},0) rectangle ({\w * 7 / 8}, {\h - 2 * \g * \h});
\draw[shortenedArrow,yshift=-7pt] ({\w * 3 / 4},0) -- node[below] {3} ({\w * 7 / 8},0);
\fill[pattern=vertical lines, pattern color=gray] ({\w * 7 / 8},0) rectangle (\w, {\h - 2 * \g * \h});
\draw[shortenedArrow,yshift=-7pt] ({\w * 7 / 8},0) -- node[below] {$\ell = 4$} (\w,0);

\draw[dashed, pattern=south west lines, pattern color=red] (0,0) rectangle ({\w * 7 / 8}, {\g*\h});

\draw[pattern=dots, pattern color=blue] ({\w / 2},{\h - 2 * \g * \h}) rectangle ({\w * 3 / 4}, \h);
\draw[->] ({\w/4},{\g*\h / 2}) -- ({\w * 5 / 8},{\h - \g*\h});

\draw[pattern=dots, pattern color=blue] ({\w * 3 / 4},{\h - 2 * \g * \h}) rectangle ({\w * 7 / 8}, \h);
\draw[->] ({\w * 5 / 8},{\g*\h / 2}) -- ({\w * 13 / 16},{\h - \g*\h});

\draw[pattern=dots, pattern color=blue] ({\w * 7 / 8},{\h - 2 * \g * \h}) rectangle (\w, {\h - \g * \h});
\draw[->] ({\w * 13 / 16},{\g*\h / 2}) -- ({\w * 15 / 16},{\h - 3*\g*\h/2});

\draw[shortenedArrow,xshift=-7pt] (0,\h) -- node[left] {\rotatebox{90}{$100\%$}} (0,{\g*\h});
\draw[shortenedArrow,xshift=-7pt] (0,0) -- node[left] {\rotatebox{90}{$\alpha$}} (0,{\g*\h});

\draw[shortenedArrow,xshift=30pt] (\w,0) -- node[left] {\rotatebox{90}{$100\%$}} (\w,{\h - \g*\h});
\draw[shortenedArrow,xshift=14pt] (\w,{\h - \g*\h}) -- node[left] {\rotatebox{90}{$\alpha$}} (\w,\h);
\draw[shortenedArrow,xshift=14pt] (\w,{\h - 2*\g*\h}) -- node[left] {\rotatebox{90}{$\alpha$}} (\w,{\h - \g*\h});

\end{tikzpicture}
}

%% file: experiments.tex
\section{Experiments}
\label{sec:experiments}

\myparagraph{Setup.} For validation we extend the experimental setup used for Cuckoo and Xor filters~\cite{FAKM:CuckooFilterReallyBetter:2014,GL:XorFilters:2020,Lemire:fastfilter:2020}, with our code available in our fork on GitHub~\cite{Dillinger:fastfilter:2021}. Timings are performed on a single-socket Intel® Xeon® D-2191 (Skylake DE) with 64GB RAM. Tests are compiled with GCC 8.4.1, using \texttt{g++ -O3 -DNDEBUG -march=\hspace{0pt}skylake-\hspace{0pt}avx512}. The Ribbon code is portable C++ using no processor intrinsics but using compiler built-ins for prefetch, count leading/trailing zero bits, and bit parity.

\begin{table}
\def\num#1{\ifnum#1<100\phantom{0}#1\else#1\fi}

 \caption{Experimental performance comparisons}
 \label{tab:performance}
 \begin{tabular}{rc|rl|rl}
  \toprule
                & Space  & \multicolumn{2}{c|}{ns/key, $n\!=\!10^6$} & \multicolumn{2}{|c}{ns/key, $n\!=\!10^8$} \\
  Configuration & ovr \% & con & query & con & query \\
  \midrule
    \multicolumn{6}{c}{$\downarrow\,\,$ FP rate around 1\%, \textbf{Ribbons} using $r=7$ $\,\,\downarrow$} \\
  \midrule
BlockedBloom\cite{LNKB:Bloom:2019}        &  52.0 &  11 & $ 14 \pm 0$ &  32 & $ \num{37} \pm 0$ \\
BlockedBloom\cite{Dillinger:RocksDBBloom} &  49.8 &  21 & $ 10 \pm 0$ &  72 & $ \num{36} \pm 0$ \\
         Cuckoo12$\dagger$ &  46.2 &  69 & $ 21 \pm 0$ & 147 & $ \num{58} \pm 0$ \\
                  Cuckoo12 &  40.3 &  91 & $ 20 \pm 0$ & 205 & $ \num{58} \pm 0$ \\
                  Morton12 &  40.5 &  87 & $ 55 \pm 3$ & 106 & $\num{125} \pm 13$ \\ %
        Xor, $r=7\ddagger$ &  23.0 & 195 & $ 21 \pm 0$ & 264 & $ \num{66} \pm 1$ \\
                Xor, $r=8$ &  23.0 & 148 & $ 15 \pm 0$ & 211 & $ \num{50} \pm 0$ \\
               Xor+, $r=8$ &  14.5 & 171 & $ 35 \pm 1$ & 299 & $\num{104} \pm 10$ \\
    \textbf{Homog.}, $w = \num{16}$ &  52.0 &  56 & $ 40 \pm 6$ & 101 & $ \num{88} \pm 6$ \\
    \textbf{Homog.}, $w = \num{32}$ &  20.6 &  58 & $ 38 \pm 6$ & 116 & $ \num{85} \pm 5$ \\
       \textbf{Standard}, $w = \num{64}$ & 14;20 &  71 & $ 42 \pm 5$ & 130 & $ \num{94} \pm 7$ \\
    \textbf{Homog.}, $w = \num{64}$ &  10.1 &  83 & $ 39 \pm 7$ & 160 & $ \num{90} \pm 6$ \\
      \textbf{Standard}, $w = \num{128}$ &   6;8 & 166 & $ 58 \pm 2$ & 235 & $\num{140} \pm 26$ \\
   \textbf{Homog.}, $w = \num{128}$ &   5.1 & 164 & $ 53 \pm 3$ & 270 & $\num{145} \pm 25$ \\
 \textbf{Balanced}, $w=32\dagger$ &  15.3 &  84 & $ 47 \pm 6$ & 278 & $\num{104} \pm 5$ \\
       \textbf{Balanced}, $w = \num{32}$ &   2.5 & 162 & $ 48 \pm 5$ & 372 & $\num{107} \pm 5$ \\
       \textbf{Balanced}, $w = \num{64}$ &   0.7 & 292 & $ 49 \pm 5$ & 516 & $\num{111} \pm 5$ \\
      \textbf{Balanced}, $w = \num{128}$ &   0.3 & 985 & $ 66 \pm 1$ & 1335 & $\num{162} \pm 29$ \\
  \midrule
  \textbf{Homog.}32, $r=7\textcolor{white}{.0}$ &  20.6 &  58 & $ 38 \pm 6$ & 116 & $ \num{85} \pm 5$ \\
  \textbf{Homog.}32, $r=7.7$ &  22.7 &  61 & $ 43 \pm 4$ & 120 & $ \num{96} \pm 7$ \\
  \textbf{Homog}.32, $r=8\textcolor{white}{.0}$ &  22.1 &  61 & $ 39 \pm 5$ & 116 & $ \num{86} \pm 9$ \\
  \midrule
    \multicolumn{6}{c}{$\downarrow\,\,$ FP rate around 10\%, \textbf{Ribbons} using $r=3$ $\,\,\downarrow$} \\
  \midrule
            Xor, $r=3\ddagger$ &  23.1 & 194 & $ 16 \pm 0$ & 264 & $ \num{55} \pm 0$ \\
    \textbf{Homog.}, $w = \num{16}$ &  34.6 &  46 & $ 23 \pm 0$ &  91 & $ \num{66} \pm 1$ \\
    \textbf{Homog.}, $w = \num{32}$ &  16.1 &  52 & $ 21 \pm 0$ & 108 & $ \num{55} \pm 1$ \\
       \textbf{Standard}, $w = \num{64}$ & 14;20 &  66 & $ 22 \pm 0$ & 124 & $ \num{62} \pm 0$ \\
    \textbf{Homog.}, $w = \num{64}$ &   8.0 &  71 & $ 21 \pm 0$ & 152 & $ \num{72} \pm 4$ \\
      \textbf{Standard}, $w = \num{128}$ &   6;8 & 145 & $ 35 \pm 0$ & 213 & $ \num{79} \pm 1$ \\
   \textbf{Homog.}, $w = \num{128}$ &   4.2 & 156 & $ 33 \pm 0$ & 266 & $ \num{79} \pm 1$ \\
       \textbf{Balanced}, $w = \num{32}$ &   2.8 & 155 & $ 29 \pm 0$ & 377 & $ \num{68} \pm 1$ \\
       \textbf{Balanced}, $w = \num{64}$ &   0.8 & 288 & $ 29 \pm 0$ & 517 & $ \num{84} \pm 11$ \\
  \midrule
    \multicolumn{6}{c}{$\downarrow\,\,$ FP rate around $2^{-11}$, \textbf{Ribbons} using $r=11$ $\,\,\downarrow$} \\
  \midrule
         Cuckoo16$\dagger$ &  36.0 &  68 & $ 20 \pm 0$ & 147 & $ \num{59} \pm 0$ \\
                  Cuckoo16 &  30.2 &  84 & $ 20 \pm 0$ & 201 & $ \num{59} \pm 1$ \\
            CuckooSemiSort &  26.6 & 146 & $ 50 \pm 0$ & 323 & $\num{143} \pm 4$ \\
                     Xor, $r=12$ &  23.1 & 175 & $ 23 \pm 0$ & 251 & $ \num{74} \pm 4$ \\
       Xor, $r=10.666\ddagger$ &  23.0 & 191 & $ 20 \pm 0$ & 284 & $ \num{61} \pm 0$ \\
          Xor+, $r=11\ddagger$ &  12.8 & 185 & $ 38 \pm 0$ & 317 & $\num{107} \pm 2$ \\
    \textbf{Homog.}, $w = \num{32}$ &  28.0 &  74 & $ 51 \pm 2$ & 129 & $\num{118} \pm 32$ \\
       \textbf{Standard}, $w = \num{64}$ & 14;20 &  78 & $ 54 \pm 3$ & 139 & $\num{138} \pm 41$ \\
    \textbf{Homog.}, $w = \num{64}$ &  12.7 &  89 & $ 51 \pm 3$ & 163 & $\num{118} \pm 26$ \\
      \textbf{Standard}, $w = \num{128}$ &   6;9 & 173 & $ 71 \pm 9$ & 247 & $\num{164} \pm 39$ \\
   \textbf{Homog.}, $w = \num{128}$ &   7.3 & 189 & $ 72 \pm 9$ & 292 & $\num{166} \pm 43$ \\
       \textbf{Balanced}, $w = \num{32}$ &   2.3 & 169 & $ 60 \pm 2$ & 394 & $\num{142} \pm 32$ \\
       \textbf{Balanced}, $w = \num{64}$ &   0.5 & 301 & $ 61 \pm 3$ & 536 & $\num{147} \pm 28$ \\
 \midrule
    \multicolumn{6}{l}{$\dagger$ Larger space allocated to improve construction time.} \\
    \multicolumn{6}{l}{$\ddagger$ Potentially unfavorable bit alignment.} \\
    \multicolumn{6}{l}{$;$ Standard Ribbon space overhead depends on $n$.} \\  \bottomrule
 \end{tabular}
\end{table}

\myparagraph{Results.} First, \cref{fig:performance-comparison} shows which approach is fastest for various space overheads and various FP rates, and query and construction times for the corresponding overall fastest approach. It uses a mix of $n ∈ \{10^6, 10^7, 10^8\}$; more details in \cref{sec:intro}. To maximize the coverage of ``BlockedBloom'' in the figure, we include an AVX2 SIMD-optimized implementation from RocksDB~\cite{Dillinger:RocksDBBloom} that sacrifices SIMD-optimized construction for enhanced configurability (any number of probes) and minimized FP rate compared to~\cite{LNKB:Bloom:2019}; all ``BlockedBloom'' in this paper use aligned 512-bit blocks. To maximize coverage of ``Bloom'' we include an obvious variant splitting probes between two independent blocks.

Second, \cref{tab:performance} shows detailed timings for some specific configurations with approximate $f ∈ \{2^{-3}, 2^{-7}, 2^{-11}\}$. Construction and query times are given in nanoseconds per key over at least $5 · 10^7$ keys, with a range of query times for different ratios of positive vs. negative queries. The timing data for $n = 10^6$ represents the core CPU time of each approach with negligible memory access overheads, while $n = 10^8$ includes memory access overheads. The reported space overhead is $\frac{r(1+ε)}{\log_2(f^{-1})}-1$ where the FP rate $f$ is measured by sampling.
Standard Ribbon has distinct overheads for the different $n$.
Cuckoo12 and Morton12 use 12 bits per cell~\cite{BJ:MortonFilters:2020,FAKM:CuckooFilterReallyBetter:2014}, for similar FP rate as $r=9$ Xor or Ribbon.

Third, \cref{tab:overheads} shows many more space overheads for Standard Ribbon, which mostly improve for smaller $n$. While we do not include timings for these smaller $n$, we can infer that Standard Ribbon, with or without smash, takes territory from other Ribbon approaches in \cref{fig:performance-comparison} \textbf{(a)} for smaller $n$.

\myparagraph{Observations.} When saving space compared to Bloom and Cuckoo (incl. Morton), we see in~\cref{tab:performance},
\begin{itemize}
    \item Ribbon can achieve much lower space overheads than the practical alternatives. $w=64$ Balanced Ribbon has less than 1\% overhead except when $r < 3$ (bottom right of \cref{fig:performance-comparison}).
    \item Ribbon mostly wins in construction time, sometimes by a large factor. Ribbon construction times are generally only higher than Xor when saving space compared to Xor. 
    \item Xor mostly wins in query time. For smaller $rw$, which is proportional to the contiguous memory loaded per query, Ribbon query times are similar to Xor. Ribbon query times increase with $r$ and/or $w$.
    \item Ribbon has a clear advantage for configurability (aside from Bloom). Xor and Cuckoo incur a measurable penalty when cells are not aligned to a friendly size, such as 4, 8, 12, or 16 bits, and have limited options for partial-bit granularity. Ribbon performance is continuous for any integer $r$, and the penalty for arbitrary fractional average $r$ is small .
    \item Ribbon can easily trade space efficiency for construction time efficiency (like Cuckoo), by increasing allocated space overhead; see $\dagger$ vs. corresponding non-$\dagger$ configurations. Xor construction time is more fixed.
\end{itemize}

A somewhat surprising result is that $w=32$ Balanced Ribbon is sometimes faster \emph{and} more space efficient than other Ribbon variants with $w=128$. We can reason that the benefits of smaller ribbon width sometimes exceed the costs associated with balancing. We consider $w=128$ Balanced Ribbon ``impractical'' because the execution time is much higher for a tiny benefit in space usage.

\myparagraph{Limitations.} A notable limitation of this test is that it tests query \emph{throughput} more than query \emph{latency}. Like executing a batch of filter queries, the test does not depend on the result of each query for what to do next. For applications executing a single filter query with unpredictable outcome, main memory latency can be 200-300ns. However, we expect latency to be similar between the competing approaches, because aside from Balanced Ribbon sharding metadata (very small, cachable) and Xor+ compression metadata (not as small), only constant size metadata is read to determine which memory to fetch to complete a query. Although the test relies on out-of-order execution between queries for memory prefetching, applications can use explicit prefetching.\footnote{RocksDB MultiGet uses batched filter queries with explicit prefetching; RocksDB Get uses single filter queries.}

To minimize sampling noise, the test machine was otherwise idle. This does not match a production environment, but we have not seen a significant difference in relative results when running the tests under load.

To match existing test code, hash seeds are not configurable. All structures are configured for high chance of construction success (roughly 99\% or more). In practice, approaches other than BlockedBloom and Homogeneous would incur some small additional overheads associated with seeding and retries.

%% file: conc-future.tex
\section{Conclusion}
Our result changes the narrative around data structures constructed with Gaussian elimination vs. with peeling. With our new algorithm and the query structure from \cite{DW:One-Block-per-Row:2019}, Gaussian elimination can be faster than peeling while also opening up better space efficiency.

On this foundation we have built Ribbon filters with the following efficiently scaling variants.
\begin{itemize}
    • The Homogeneous Ribbon filter is simple and has a construction algorithm that never fails. We provide a full analysis.
    • The Balanced Ribbon filter leverages the on-the-fly and incremental construction algorithm in an experimental load-balancing scheme that further reduces space overhead.
\end{itemize}

As with Bloom filters, the true practicality of Ribbon also comes from being able to configure it for or adapt it to the application, including dynamic conditions. This includes using any amount of space to efficiently represent any number of keys, even by shrinking the structure after construction. Ribbon filters replicate the smooth FP-rate-for-space configurability of Bloom filters and extend that with configurability between space efficiency and time efficiency.\\

\noindent \textbf{Future work} could pursue the following goals.

\begin{itemize}
  • Deepen the theoretical understanding of Ribbon filters to better guide parameter choices in practice.
  • Manage the variance of FP rate of Homogeneous Ribbon filters at small scales.
  • Further explore the design space surrounding Balanced Ribbon for approaches that are better and/or easier to analyze. %
  • Create a SIMD-optimized implementation of Ribbon queries, perhaps using AVX512 POPCNT and/or 8-bit ICML rather than $w$-bit ICML.
  • Explore how competitive Ribbon can be as a data structure for static functions, including cases where the number $r$ of solution columns is quite high, e.g. $r ≥ 32$.
\end{itemize}